\documentclass[aps,pra,twocolumn,showpacs,superscriptaddress,groupedaddress]{revtex4-2}
\usepackage[utf8]{inputenc}
\usepackage{amsmath, amsthm, amssymb, amsfonts,enumitem}
\usepackage[makeroom]{cancel}
\usepackage{graphicx}
\usepackage{bm}
\usepackage{dsfont}
\usepackage{mathtools}
\usepackage{booktabs}
\usepackage{thm-restate}
\usepackage[colorlinks=true, linkcolor=blue, urlcolor=blue,citecolor=blue]{hyperref}
\usepackage{orcidlink}
\DeclareMathOperator{\diag}{diag}

\DeclareMathOperator{\tr}{tr}

\DeclareMathOperator{\rank}{rank}

\newcommand{\braket}[2]{\mathinner{\langle #1|#2\rangle}}
\newcommand{\dyad}[1]{| #1\rangle \langle #1|}

\newcommand{\one}[0]{\mathds{1}}

\newcommand{\R}{\mathds{R}}
\newcommand{\C}{\mathds{C}}

\newcommand{\cormax}{c_{\max}}
\newcommand{\cormin}{c_{\min}}

\DeclareUnicodeCharacter{202C}{\^{i}}

\makeatletter
\newcommand{\vast}{\bBigg@{4}}
\newcommand{\Vast}{\bBigg@{5}}
\makeatother
\newtheorem{theorem}    {Theorem}

\newtheorem{proposition}[theorem]{Proposition}
\newtheorem{observation}[theorem]{Observation}

\usepackage{soul,xcolor}

\begin{document}

\title{
Communication inequalities: robust quantum advantage, measurement testing and dimension witnessing with minimal setups
}
\author{Albert Rico$^{1}$$^{\orcidlink{0000-0001-8211-499X}}$, Mani Zartab$^2$$^{\orcidlink{0000-0003-2437-0906}}$, Javier Fernández$^2$$^{\orcidlink{0009-0006-5765-0129}}$, John Calsamiglia$^2$$^{~\orcidlink{0000-0003-1735-1360}}$, Some Sankar Bhattacharya$^2$$^{\orcidlink{0000-0001-6464-5068}}$}
\affiliation{$^1$Naturwissenschaftlich-Technische Fakult\"{a}t, Universit\"{a}t Siegen, Walter-Flex-Stra\ss e 3, 57068 Siegen, Germany}
\affiliation{$^2$ Physics department, Universitat Autònoma de Barcelona, ES-08193 Bellaterra (Barcelona), Spain}

\date{\today}

\begin{abstract}
Communication scenarios (CS) are arguably the simplest setting in which quantum behaviors can be observed. Yet, in contrast to Bell scenarios or prepare-measure scenarios with shared randomness, no general method is known to separate classical and quantum statistics. Here we close this gap by deriving dimensional inequalities to separate classical and quantum statistics in CS. These depend only on four parameters of the input-output statistics, corresponding to correlated and anticorrelated events. Under dimensional assumptions, saturating one of the inequalities certifies SIC-POVMs with analytical noise robustness. The inequalities also yield a dimension witness for both correlated and anti-correlated statistics, without the need of measurement incompatibility.
\end{abstract}

\maketitle

\section{Introduction}
Quantum features have led to information-processing protocols that outperform classical ones in tasks such as computation~\cite{Shor1995PolynomialTimeAF,Grover1996Search}, communication~\cite{mattle1996dense}, and metrology~\cite{zhou2018HLimitNature}. These advantages arise because quantum systems allow for phenomena such as superposition, entanglement, and incompatibility of measurements, which are absent in classical theories.
A central challenge is therefore to identify genuinely quantum behavior from experimentally accessible data under physically motivated assumptions~\cite{Guhne_2009EntDet,brunner2014bell,Contextuality2022Review}. Besides their foundational relevance, such certification methods are essential for validating quantum devices and for guaranteeing the origin of the observed advantage.

The most common strategy to display a quantum-classical gap is based on establishing \emph{inequalities} that classical correlations of the system must satisfy, whose violation identifies quantum behavior. These provide a continuous separation between quantum and classical correlations and allow for possibilities like noise-tolerant quantum certification and quantitative quantum advantage, enabling tests of fundamental quantum properties such as dimensionality~\cite{Brunner2008DimWItTestingnonloc,Pal2008DimWitHighDimNonLoc,Vertesi2010DimWitLoopHoleNonLoc,Hendrych_2012DimWitnessExperiment}. In some cases it is possible to self-test~\cite{Supic2020Self-testReview}, namely to assess a unique quantum mechanical description (e.g. maximal entanglement) of the underlying system in the event of maximal quantum violation.
The earliest examples are Bell inequalities to certify nonlocality of space-like correlations without communication~\cite{brunner2014bell}. Later, a similar formalism has been applied to time-like correlations between preparations and measurements~\cite{Emary_2014Legggett-GargIneq,Fritz_2010TemporalCHSH,Egelhaaf_2025CertIncPrepMeasBell}, designed for communication setups with bounded dimensionality.

\begin{figure}[tbp]
    \centering
    \includegraphics[width=\linewidth]{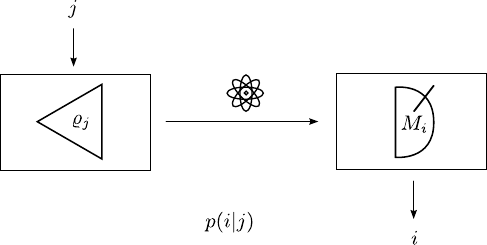}
    \caption{{\bf Minimal communication scenario (CS).} A quantum state $\varrho_j$ is prepared, and then measured with a quantum measurement $\mathcal{M}=\{M_i\}$, without other resources between the devices like shared randomness or entanglement. We analyze the probabilities $p(i|j)$ of obtaining outcome $i$ given preparation $j$. This protocol has been considered in setups displaying quantum advantage~\cite{RandFree_23_Some,Heino24SimpleUnboundQAdvtg,Ding2024quantum}, and involves fewer resources than more standard setups allowing for an input label choosing the measurement device~\cite{Drotos2024SelfTestSICpovms,Miklin2021universalSelfTestSIC,tavakoli2020self,Tavakoli2018SelfTestPM}.}
    \label{fig:prepmeas}
\end{figure}

For near-term quantum communication, distributing quantum resources like entanglement or measurement incompatibility is still a major challenge due to experimental noise. This motivates considering the simplest possible communication scenarios (CS) minimizing the resources involved: Alice receives an input, sends a message to Bob, and Bob reports an output without using any additional input nor shared resources with Alice (Fig.~\ref{fig:prepmeas})~\cite{RandFree_23_Some,Heino24SimpleUnboundQAdvtg}. The correlations obtained in this setup are known as {\em communication matrices (CM)}, and have been shown to be a promising tool to show quantum advantage: contrary to asymptotic setups~\cite{holevo1973bounds}, these single-shot prepare-measure protocols accommodate large dimensional gaps between quantum and classical strategies to reproduce target correlations~\cite{RandFree_23_Some,Heino24SimpleUnboundQAdvtg}, and have gained increasing interest~\cite{puliyil2025SDVIcomMats,Heinosaari2024maxElsCommuns,Ding2024quantum} due to their simplicity. The main difference with respect to frameworks such as nonlocality~\cite{brunner2014bell}, steering~\cite{Uola2020ResourceExclusion} or contextuality~\cite{Contextuality2022Review}, is that quantum advantage in CM is based solely on detecting the ability of preparing and measuring sets of quantum states with superposition~\cite{Designolle2021CoherenceSet}. However, in contrast to other prepare-measure schemes using shared randomness~\cite{VanHimbeeck2017SDVIPMwithRandomness,brask2026IneqPM} or external inputs for Bob~\cite{tavakoli2020self,Tavakoli2018SelfTestPM,Gois2021IneqPM}, the study of quantum advantage in these minimalistic setups is at the moment mostly based on identifying exemplary instances of CMs.
The main challenge is that, in the absence of shared randomness, the sets of CMs achievable with fixed classical dimension are generally non-convex~\cite{JdVicente2017SRandDVIdimWit}. Consequently, standard convex-separation methods do not provide a systematic characterization of these dimension-constrained correlations~\cite{brask2026IneqPM}.

In this work we derive  inequalities to distinguish families of CM displaying quantum advantage, beyond isolated examples (Theorem~\ref{thm:dCboundGeneral}, Corollary~\ref{thm:dCbound} and Proposition~\ref{prop:dQupper}). These provide a continuous formulation for prepare-measure quantum advantage with intrinsic robustness, and allow us to derive properties of the gap between classical and quantum strategies: {\em First}, saturation of the inequalities identifies a one-parameter family of CMs realizable with quantum dimension $d_Q=\sqrt n$ but requiring classical dimension $d_C=n=d_Q^2$, which contains the most noise-robust quadratic dimension separation (Eqs.~\eqref{eq:1-paramPdist} and~\eqref{eq:RobustQuadAdv}, Lemma~\ref{lem:detbound}).
{\em Second}, by saturating one of the inequalities, the statistics certify symmetric informationally complete quantum measurements (SIC-POVMs) under dimensional assumptions (Theorem~\ref{thm:selftest}). This significantly reduces the resources needed for this purpose with respect to standard prepare-measure scenario~\cite{tavakoli2020self,Tavakoli2018SelfTestPM,Drotos2024SelfTestSICpovms,Miklin2021universalSelfTestSIC,Supic2020Self-testReview}. {\em Third,} we provide robustness bounds for this certification, including leakage outside the assumed Hilbert space (Propositions~\ref{prop:continuity} and~\ref{prop:leakage}, and Theorem~\ref{thm:robustSIC}). {\em Fourth}, the same inequalities lead to a witness for the Hilbert space dimension of the underlying quantum system that applies to both correlated and anti-correlated statistics (Observation~\ref{obs:DimWit}), without the need of shared entanglement or inputs for Bob's device~\cite{Pal2008DimWitHighDimNonLoc,Brunner2008DimWItTestingnonloc,Vertesi2010DimWitLoopHoleNonLoc,Hendrych_2012DimWitnessExperiment}. These results are summarized in Figure~\ref{fig:Sets-dQ-dC}.

Beyond identifying full regions in the set of CM, the presented inequalities also allow one to test quantum advantage of specific correlations with complexity $\mathcal{O}(n^2)$, in contrast to $\mathcal{O}(n^3)$ for standard rank estimation, or approximately $\mathcal{O}(n^{2.37})$ with the best known algorithms~\cite{coppersmith1987RankPoly2,Alman_2024RankPoly2}.

\begin{figure}[tbp]
    \centering
    \includegraphics[width=\linewidth]{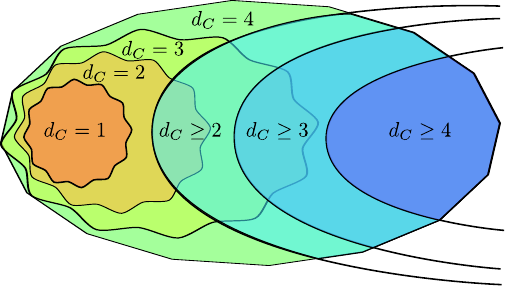}
    \caption{{\bf Inequalities for the classical dimension of CMs.} The sets of communication matrices (CMs) $P$ that can be obtained with classical dimension $d_C$ (orange, yellow and green) are not convex. Our main goal is to derive inequalities characterizing properties of CMs, for instance detecting sets of CMs requiring a certain $d_C$ to be simulated classically (blue shaded regions). These are in general nonlinear due to the non-convexity of the sets, in contrast to standard approaches for convex sets such as entanglement witnesses.
    }
    \label{fig:SetsIneq}
\end{figure}

\section{Setup}

We consider the CS setup: first, Alice prepares a message $j$. Then she sends it to Bob, who performs a measurement with outcome $i$. Finally, they evaluate the conditional probability $p(i|j)$ of obtaining the $i$'th measurement outcome given that the $j$'th preparation was done~\cite{Heino24SimpleUnboundQAdvtg,Ding2024quantum,puliyil2025SDVIcomMats,Heinosaari2024maxElsCommuns} (Fig.~\ref{fig:prepmeas}). In any physical theory describing the preparation and measurement devices, we consider the operational dimension $d$, namely the number of states that can be distinguished with probability 1 (i.e., single-shot). For instance, if $p(i|j)$ is reproduced classically, a $d_C$-dimensional simplicial alphabet is used to encode (prepare) and read (measure) the message. If $p(i|j)$ is reproduced with quantum devices, then a quantum state acting on a $d_Q$-dimensional Hilbert space is prepared and measured. Thus $\log d_C$ and $\log d_Q$ are respectively the classical and quantum information capacities, i.e., the bandwidths of the channel through which Alice and Bob communicate. In turn, $C_d$ and $Q_d$ denote the sets of classical and quantum CMs that can be obtained with operational dimension $d$. Throughout, we assume that the preparation and measurement devices are uncorrelated, i.e., they share no randomness, including that induced by a common drift of the devices~\cite{VanHimbeeck2017SDVIPMwithRandomness,JdVicente2017SRandDVIdimWit}.

When comparing quantum and classical strategies to reproduce $p(i|j)$, quantum advantage appears if the dimension $d_C$ required to generate the correlations $p(i|j)$ is strictly larger than the dimension $d_Q$ required through the Born rule, $p(i|j) = \tr(\varrho_j M_i)$ with quantum states $\varrho_j$ and a measurement with elements $M_i$. The classical dimension is equal to the so-called {\em nonnegative rank} of the communication matrix $P$ with entries $P_{ij}=p(i|j)$~\cite{jain2013efficient,cohen1993NNrank,fiorini2012NNrankdC,gillis2012geometricNNrank}. That is, the smallest number of rank-one nonnegative matrices needed to decompose $P$ in convex combination. While this quantity is in general difficult to compute~\cite{Fawzi_2014LBnnegRank,shitov2017complexityNNRank,vavasis2010complexityNNrank,shitov2014upperBoundNNrank}, the standard rank of $P$ serves as a lower bound, namely $\rank(P)\leq d_C$~\cite{JdVicente2017SRandDVIdimWit}. The quantum dimension $d_Q$ is the smallest possible size of $\varrho_j$ and $M_i$ such that $p(i|j)=\tr(\varrho_jM_i)$, and it corresponds to the so-called {\em positive-semidefinite rank} of $P$~\cite{fawzi2015psdRank}. In turn, quantum advantage is verified if $d_Q<\rank(P)$.

The bounds above, however, do not directly provide a clear understanding of what physical strategies require larger dimension to be simulated by classical systems. Instead, exemplary instances of the type of Fig.~\ref{fig:prepmeas} have been shown to provide significant gaps $d_C>d_Q$ between classical and quantum theory~\cite{RandFree_23_Some,Heino24SimpleUnboundQAdvtg,Chakraborty2025AdvQubitComPM,doolittle2025OpFrQComNettsPM}. This existing literature suggests that anti-correlated probability distributions are promising candidates to find examples of such quantum advantage, but to our knowledge no formal proof has been found. Moreover, exact correlations are not reproducible in experiments, since practical devices are inherently noisy. Robustness is addressed with the eigenvalues (or singular values) of a CM (Fig.~\ref{fig:Sets-dQ-dC}), which is specific of each case~\cite{ExPaper}. To close this gap, in this work we derive continuous inequalities for such dimensions depending on few parameters.

\section{Inequalities for communication matrices}
Which CMs cannot be reproduced with small classical systems? Here we answer this question with lower bounds on $d_C$ using only the largest and smallest probabilities in the diagonal and off-diagonal of $P$. This gives a continuous notion of which correlated and anti-correlated strategies lead to quantum advantage.

For convenience, we define the following
\begin{align}
    c_{\min}:=\min_i P_{ii} \quad & c_{\max}:=\max_i P_{ii},\label{eq:c_minmax}\\
    a_{\min}:=\min_{i\neq j}P_{ij}\quad & a_{\max}:=\max_{i\neq j}P_{ij}.\label{eq:a_minmax}
\end{align}

\begin{figure}
    \centering
    \includegraphics[width=1.05\linewidth]{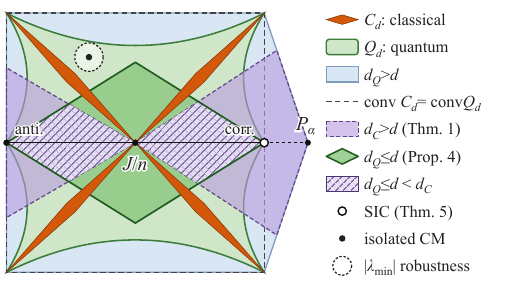}
    \caption{{\bf Sets of classical and quantum CMs}, and inequalities derived in this work. The convex hulls ${\rm conv}$ of $C_d$ and $Q_d$ coincide, and are obtained when shared randomness is allowed~\cite{Frenkel_2015ComMats}. We find upper and lower bounds for $d_C$ and $d_Q$, particularly tailored to  correlations (corr.) and anticorrelations (anti.). Hence we detect quantum advantage in the whole region $d_Q\leq d<d_C$. Saturation of Eq.~\eqref{eq:dCboundDisc} certifies SIC-POVMs, and we denote $J$ with $J_{ij}=1$. These inequalities provide a systematic characterization from few parameters, in contrast to isolated examples, whose robustness is addressed via rank stability.}
    \label{fig:Sets-dQ-dC}
\end{figure}

\begin{restatable}[CM inequalities]{theorem}{dCboundGeneral}\label{thm:dCboundGeneral}
Let $P$ be a CM of size $n$, with $P_{ij}=p(i|j)$. The dimension $d_C$ needed to simulate $P$ classically is lower bounded by $d_C\geq1+\Gamma$, where
\begin{equation}\label{eq:UnifiedBound}
    \Gamma= \frac{\max\{n c_{\min}-1,1-nc_{\max},\;0\}}{1 - (n-2)a_{\min} - 2\min\{a_{\min},c_{\min}\}\,}\,.
\end{equation}
\end{restatable}
The proof is given in Appendix~\ref{app:ProofGenIneq}. It follows that quantum advantage is certified whenever the target statistics $P$ are generated with a quantum system of dimension $d_Q<1+\Gamma$, since no classical system of the same dimension can reproduce them. The inequalities derived are represented in Fig.~\ref{fig:Sets-dQ-dC}.
This inequality leads us to identify the following special families of CMs, tailored for Eqs.~\eqref{eq:dCboundDisc} and~\eqref{eq:dCboundExcl} respectively:
\begin{enumerate}
    \item {\em Correlated strategies}, where correlated events are more likely to occur than anti-correlated events, namely $c_{\min}\geq a_{\max}$ and therefore $p(i|i)\geq p(i|j)$ for all $i,j$.
    \item {\em Anti-correlated strategies}, where correlated events are less likely to occur than anti-correlated events, namely $c_{\max}\leq a_{\min}$ and therefore $p(i|i)\leq p(i|j)$ for all $i,j$.
\end{enumerate}
To the best of our knowledge, existing literature on CS has solely considered anticorrelated and heuristic strategies~\cite{RandFree_23_Some,Heino24SimpleUnboundQAdvtg,ExPaper}. In this context, Theorem~\ref{thm:dCboundGeneral} allows us to characterize a broader set systematically, and leads us to two conclusions: first, that anticorrelated strategies systematically require a large classical dimension, which explains their known quantum advantage; and second, that correlated strategies display as well a quantum-classical gap. This is formulated as follows.

\begin{restatable}{corollary}{dCboundCorollary}
\label{thm:dCbound}
Let $P$ be an $n\times n$ column-stochastic communication matrix.
If $P$ is correlated, then
\begin{equation}\label{eq:dCboundDisc}
    d_C\geq 1+\frac{n c_{\min}-1}{1-n a_{\min}} .
\end{equation}
If $P$ is anti-correlated, then
\begin{align}\label{eq:dCboundExcl}
    d_C &\geq 1+\frac{1-n c_{\max}}{1-2c_{\min}-(n-2)a_{\min}}
\end{align}
The  bounds are understood outside the trivial uniform case $P=J/n$, for which $d_C=1$.
\end{restatable}

The proof is given in Appendix~\ref{app:ProofIneq}. Operationally, the inequality~\eqref{eq:dCboundDisc} is saturated by state discrimination protocols, relevant for practical communication~\cite{Barnett2009QSDisc,Bae2015ASDiscApps}, and Eq.~\eqref{eq:dCboundExcl} by state exclusion protocols, suitable for noise simulation~\cite{Brun2002QSExcl,Bandyopadhyay2014QSExcl}, in a noise-tolerant way. In their quantum realization, the goal is respectively to ensure or exclude that a state preparation $\varrho_j$ with label $j$ was performed, by means of a single quantum measurement $\{M_i\}$. For this purpose, we design the measurement operators to maximize (minimize) $p(i|j=i)=\tr(\varrho_{j=i}M_i)$, and minimize (maximize) the probabilities $p(i|j\neq i)=\tr(\varrho_{j\neq i}M_i)$.

Geometrically, the idea behind the inequalities~\eqref{eq:dCboundDisc} and~\eqref{eq:dCboundExcl} is analogous to linear witnesses separating regions of convex sets (e.g. of entangled states or nonlocal correlations)~\cite{Guhne_2009EntDet,brunner2014bell}. The main difference is that here the sets of correlations are not convex due to the lack of shared randomness and similar resources (see Fig.~\ref{fig:SetsIneq}), and therefore standard methods such as the Hahn-Banach theorem~\cite{Hahn1927,Banach1929} do not immediately apply. Due to this reason, existing methods~\cite{RandFree_23_Some,Heino24SimpleUnboundQAdvtg,ExPaper,Chakraborty2025AdvQubitComPM} are limited to exemplary points in the space of correlations, and the most systematic existing inequality was still the rank of the overall CM~\cite{JdVicente2017SRandDVIdimWit}, which indeed lower bounds the nonnegative rank~\cite{Heino24SimpleUnboundQAdvtg}, but is tailored to specific instances. In this sense, the inequalities above resolve this challenge, by providing a nonlinear cut on the space of probabilities $P$ into regions of correlations requiring dimensionality $d_C$ to be reproduced classically (Fig.~\ref{fig:SetsIneq}). Therefore, to our knowledge, the inequalities above provide the first inequality-based continuous separation between classical and quantum CM, and an intrinsic notion of robustness for quantum advantage in operational dimension. Moreover, they are analytical, depend only on the diagonal and off-diagonal extrema, and can be used to certify SIC-POVMs, as we show in Section~\ref{sec: SIC-POVM self-test}.

Since the inequalities introduced only use extremal entries, the bounds can be sensitive to a few outlying probabilities, such as a single small off-diagonal entry. This can be circumvented with two simple observations. First, relabelling the outcomes permutes the rows of $P$ without changing its rank, so Theorem~\ref{thm:dCboundGeneral} can be applied to the labelling that best aligns the largest probabilities with the diagonal. Second, the rank of a principal submatrix of $P$ cannot exceed that of $P$, and is invariant under rescaling its columns. Theorem~\ref{thm:dCboundGeneral} therefore also applies to any principal submatrix with renormalized columns, which allows one to discard outlying entries while still lower bounding $d_C$.

\section{Noise-tolerant quadratic quantum advantage}

Our next goal is to identify the CMs that saturate the inequalities above. To this end, first notice that any CM admitting a quantum realization $p(i|j)=\tr(\varrho_jM_i)$ on a Hilbert space of dimension $d_Q$ satisfies $\rank(P)\leq d_Q^2$~\cite{JdVicente2017SRandDVIdimWit}, since a measurement on $\C^{d_Q}$ contains at most $d_Q^2$ linearly independent effects. Therefore, saturating this bound leads to quadratic quantum advantage.  More precisely, a CM $P$ of size $n=d_Q^2$ certifies quadratic advantage whenever (1) it admits a quantum realization in dimension $d_Q=\sqrt{n}$, and (2) its smallest eigenvalue in absolute value, $\lambda_{\min}$, is nonzero, so that $\rank(P)=n$. This requires $P$ being full rank, and therefore it is necessary that both the states and the measurements are tomographically complete (that is, linearly independent operators).

In any practical implementation the target is reproduced only up to an experimental CM $P'$, deviating from $P$ due to systematic and statistical inaccuracies in the devices. The smallest singular value of $P'$ differs from that of $P$ by at most $\|P-P'\|_2$ (Appendix~\ref{app:RankStabilityPalpha}). For the CMs singled out below by Lemma~\ref{lem:maxSaturationPalpha}, which are symmetric, the latter equals $|\lambda_{\min}|$. Hence $|\lambda_{\min}|$ quantifies the amount of noise that the advantage tolerates before the rank of $P$ drops. For this reason we focus on such CMs that maximize $|\lambda_{\min}|$, namely satisfying $|\lambda_{\min}|=|\lambda_{2}|$ (by ordering the eigenvalues non increasingly, and considering that the largest eigenvalue of any stochastic matrix is 1). This leads us to the following family of CMs~\cite{Heinosaari2024maxElsCommuns},

\begin{equation}\label{eq:1-paramPdist}
    P_\alpha=\alpha\one + \frac{1-\alpha}{n-1}(J-\one),
\end{equation}
i.e., $p(i|i)=\alpha$ and $p(i|j)=\tfrac{1-\alpha}{n-1}$, with $\one_{ij}=\delta_{ij}$ the identity and $J_{ij}=1$ the all-ones matrix. Indeed, the smallest eigenvalue is $\lambda_{\min}=|n\alpha-1|/(n-1)$ with multiplicity $n-1$, nonzero for every $\alpha\neq 1/n$, which coincides with its smallest singular value since $P_\alpha$ is symmetric.

Any experimental CM $P'$ whose deviation from $P_\alpha$ obeys
\begin{equation}\label{eq:RobustQuadAdv}
    \|P_\alpha-P'\|_2<\frac{|n\alpha - 1|}{n-1},
\end{equation}
will satisfy $d_C=n$ (see Appendix~\ref{app:RankStabilityPalpha}), where $\|\cdot\|_2$ denotes the spectral norm (largest singular value). If every entry of $P'$ deviates from $P_\alpha$ by at most $\varepsilon$, then $\|P_\alpha-P'\|_2\leq n\varepsilon$, so that $\varepsilon<|n\alpha-1|/(n(n-1))$ suffices. Alternatively, since $d_C$ is an integer, Theorem~\ref{thm:dCboundGeneral} certifies $d_C=n$ directly from the three extremal entries of $P'$ whenever $\varepsilon<|n\alpha-1|/(n(n-1)^2)$ (Appendix~\ref{app:RankStabilityPalpha}). This condition is more conservative by a factor $n-1$, but it requires no spectral information. In this regime, quadratic quantum advantage is guaranteed whenever $P'$ is produced with a $d_Q=\sqrt{n}$-dimensional quantum system.

In fact, the family of CM in Eq.~\eqref{eq:1-paramPdist} is naturally selected by Theorem~\ref{thm:dCboundGeneral}:
\begin{restatable}[Maximal saturation selects $P_\alpha$]{lemma}{maxSaturationPalpha}\label{lem:maxSaturationPalpha}
Let $P$ be an $n\times n$ column-stochastic CM with $n\geq3$. If the unified bound~\eqref{eq:UnifiedBound}, with positive denominator, attains the maximal value $1+\Gamma=n$, then $P=P_\alpha$
with $\alpha\neq1/n$. Conversely, every $P_\alpha$ with $\alpha\neq1/n$ saturates~\eqref{eq:UnifiedBound} at $d_C=n$.
\end{restatable}
The proof is given in Appendix~\ref{app:select_Palpha}. Thus, the family $P_\alpha$ is singled out rigorously by our inequalities: it contains the only CMs for which Eq.~\eqref{eq:UnifiedBound} attains the maximal value $1+\Gamma=n$. This family was introduced in Ref.~\cite{Heinosaari_2020ComMats} (there denoted $D_{n,1-\alpha}$), where its rank and its realizability in dimension $\sqrt{n}$ were analyzed.

In particular, $P_\alpha$ with $\alpha>1/n$ ($\alpha<1/n$) is correlated (anti-correlated) and saturates Eq.~\eqref{eq:dCboundDisc} (Eq.~\eqref{eq:dCboundExcl}). As $\alpha$ varies, it interpolates between the two regimes. For $1/n\leq\alpha\leq 1$ the CM is correlated, with the upper edge $\alpha=1$ giving the maximal identification probability of every signal. For $0\leq\alpha\leq 1/n$ it is anti-correlated, with the lower edge $\alpha=0$ giving the optimal strategy when the receiver must avoid reporting the same index as the sender (exclusion). The two regimes meet at $\alpha=1/n$, the unique point where $P_\alpha=J/n$ has rank one and hence $d_C=1$.

\begin{figure}
    \centering
    \includegraphics[width=\linewidth]{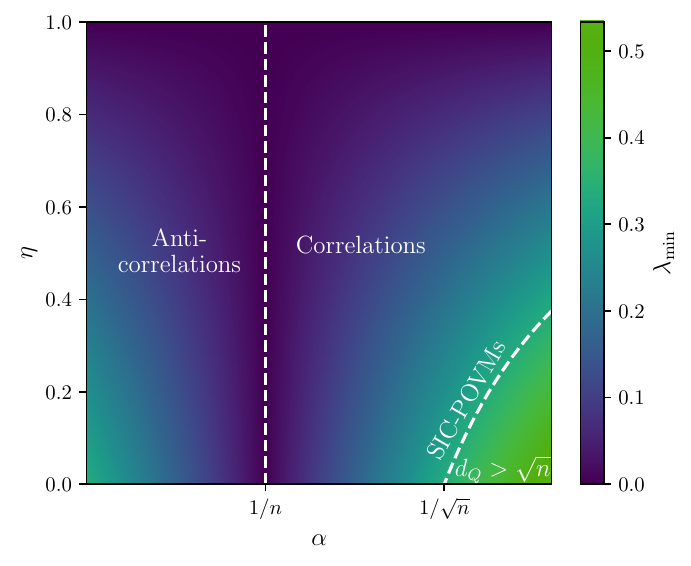}
    \caption{{\bf Quadratic quantum advantage and noise tolerance.} Noisy instances $P_{\alpha,\eta}=(1-\eta)P_\alpha+\eta J/n$ of Eq.~\eqref{eq:1-paramPdist}. The color indicates the smallest eigenvalue $\lambda_{\min}$ in absolute value for $n=4$, which indicates the noise tolerance up to which quadratic advantage holds through $d_C=\rank(P_{\alpha,\eta})=n$. The vertical dashed line ($\alpha=1/n$, where $P_\alpha=J/n$ has rank one) separates anti-correlations from correlations, both achievable with $d_Q=\sqrt{n}$-dimensional quantum devices. The curved dashed line separates the region $\alpha> 1/\sqrt{n}$, where $P_\alpha$ can only be reproduced with $d_Q>\sqrt{n}$ or possibly with a more general probability theory (GPT) in operational dimension $\sqrt{n}$. Since $P_{\alpha,\eta}=P_{\alpha'}$ with $\alpha'=(1-\eta)\alpha+\eta/n$, white noise only moves the CM along the family, and $|\lambda_{\min}|=(1-\eta)|n\alpha-1|/(n-1)$: the vertical axis shows the amount of depolarizing noise tolerated at each $\alpha$, and the curved line is $\alpha'=1/\sqrt{n}$.}
    \label{fig:Palpha}
\end{figure}

Classically, every other value requires the maximal dimension $d_C=n$. Quantum mechanically, a system of dimension $d_Q=\sqrt{n}$ can only reach $\alpha\leq 1/d_Q=1/\sqrt{n}$, since $n\alpha=\tr(P)\leq d_Q$~\cite{Heinosaari_2020ComMats}, and it reaches the whole range $0\leq\alpha\leq 1/\sqrt{n}$ whenever a SIC-POVM exists in dimension $\sqrt{n}$. Operationally, the edges correspond to optimal state exclusion at $\alpha=0$ and optimal state discrimination at $\alpha=1/\sqrt{n}$. To achieve larger values $\alpha>1/\sqrt{n}$, one needs a larger dimensional quantum system, $d_Q>\sqrt{n}$ (curved dashed line in Fig.~\ref{fig:Palpha}). This provides a formal explanation for the observation that anti-correlations display quantum advantage~\cite{Heino24SimpleUnboundQAdvtg,RandFree_23_Some}. It also shows that the same dimensional gap $d_Q=\sqrt{n}<n=d_C$ occurs for correlated CMs at the discrimination edge $\alpha=1/\sqrt{n}$, consistently with the analysis of this family in~\cite{Heinosaari_2020ComMats}. Moreover, the correlated edge is more noise tolerant: $|\lambda_{\min}|=1/(\sqrt{n}+1)$ at $\alpha=1/\sqrt{n}$, compared to $1/(n-1)$ at $\alpha=0$. Both coincide for qubits, and correlations are increasingly favoured for $d_Q\geq3$.

Intermediate values $0<\alpha<1/\sqrt{n}$ are reached, e.g., by depolarizing the SIC states, $\varrho_j=(1-t)\Pi_j+t(\one-\Pi_j)/(d_Q-1)$ with $t\in[0,1]$, and admit multiple inequivalent realizations, since the noise can be placed in the states or in the measurement; only the two edges are therefore candidates for certification (Section~\ref{sec: SIC-POVM self-test}). Moreover, the discrimination edge is the most noise-tolerant CM of its size among all those realizable in dimension $d_Q$:
\begin{restatable}[Maximal noise tolerance of the discrimination edge]{lemma}{detBound}\label{lem:detbound}
Let $n=d^2$ and let $P$ be an $n\times n$ CM generated by states and a measurement on $\C^d$. Then $|\det P|\leq(d+1)^{-(n-1)}$. Consequently, the eigenvalue and the singular value of $P$ of smallest modulus satisfy $|\lambda_{\min}|\leq 1/(d+1)$ and $\sigma_{\min}\leq1/(d+1)$, with equality for $P_\alpha$ with $\alpha=1/d$.
\end{restatable}
The proof is given in Appendix~\ref{app:RankStabilityPalpha}. Together with Eq.~\eqref{eq:RobustQuadAdv}, Lemma~\ref{lem:detbound} shows that no $d_Q$-dimensional realization of a quadratic advantage tolerates more noise, in the sense of rank stability, than the SIC discrimination strategy.

\section{Quantum advantage from four parameters}
With Theorem~\ref{thm:dCboundGeneral} we have lower bounded $d_C$. In order to obtain quantum advantage, we further require that the same statistics can be produced with a quantum system of a smaller dimension. Here we show that this can be decided from the statistics alone, without knowledge of the underlying realization, using only the four extremal entries of $P$ given in Eqs.~\eqref{eq:c_minmax} and~\eqref{eq:a_minmax}. For that, let us recall that the dimension $d_Q$ is the (complex) positive-semidefinite rank of $P$~\cite{fawzi2015psdRank,jain2013efficient}, for which upper bounds are typically obtained from explicit factorizations, e.g., via Hadamard square roots~\cite{LeeWeiDeWolf2017PSDrank}. For CMs, such a factorization can be constructed explicitly:

\begin{restatable}[Four-parameter bound on the quantum dimension]{proposition}{dQupper}\label{prop:dQupper}
Let $P$ be an $n\times n$ CM with $n\geq2$, and let $\delta_c:=\max\{|c_{\max}-\tfrac1n|,|c_{\min}-\tfrac1n|\}$ and $\delta_a:=\max\{|a_{\max}-\tfrac1n|,|a_{\min}-\tfrac1n|\}$. If
\begin{equation}\label{eq:dQupper}
    \delta_c^2+(n-1)\,\delta_a^2\leq\frac{1}{(\lceil\sqrt{n}\rceil-1)^2\,n(n-1)},
\end{equation}
then $P$ admits a quantum realization in dimension $d_Q\leq\lceil\sqrt{n}\rceil$.
\end{restatable}

The quantum advantage certified by combining this bound and Theorem~\ref{thm:dCboundGeneral} is depicted in Fig.~\ref{fig:Sets-dQ-dC}.
The proof, given in Appendix~\ref{app:dQupper}, is based on constructing specific states and measurements in dimension $\lceil\sqrt{n}\rceil$. Combined with Theorem~\ref{thm:dCboundGeneral}, this yields a certificate of quantum advantage that depends only on the four parameters of Eqs.~\eqref{eq:c_minmax} and~\eqref{eq:a_minmax}: if Eq.~\eqref{eq:dQupper} holds and $1+\Gamma>\lceil\sqrt{n}\rceil$, one has
\begin{equation}
    d_Q\leq\lceil\sqrt{n}\rceil<1+\Gamma\leq d_C .
\end{equation}
Such CMs exist for every $n\geq3$. For instance, $P_\alpha$ with $0<|\alpha-1/n|<1/[n(\lceil\sqrt{n}\rceil-1)]$ satisfies Eq.~\eqref{eq:dQupper} and has $1+\Gamma=n$. Any sufficiently small perturbation of it still satisfies Eq.~\eqref{eq:dQupper} and $1+\Gamma>n-1$, which suffices since $d_C$ and $d_Q^2$ are integers. Hence quantum advantage in CMs is not restricted to isolated points, but holds in larger continuous regions. For $n=d_Q^2$ and in dimensions where SIC-POVMs exist, measuring with the SIC-POVM yields a complementary region $(d_Q-1)a_{\min}+\min\{a_{\min},c_{\min}\}\geq1/(d_Q+1)$, which is saturated by both optimal discrimination and optimal exclusion (Appendix~\ref{app:dQupper}).

\section{Robust semi-device-independent certification of SIC-POVMs}\label{sec: SIC-POVM self-test}

In the best-case scenario, saturation of correlation inequalities may allow for self-testing~\cite{Supic2020Self-testReview}, namely identifying a unique maximally resourceful quantum strategy up to isometries. In space-like and time-like correlations, one can certify maximal entanglement in shared states~\cite{brunner2014bell} and particular quantum measurements~\cite{Tavakoli2018SelfTestPM,tavakoli2020self,Drotos2024SelfTestSICpovms}, respectively. However, it remains unclear whether a similar certification can be done in CS, where neither shared resources nor measurement incompatibility are available. A first step was made in~\cite{puliyil2025SDVIcomMats} for channel identification from specific instances of CMs. Here we show that, under the dimensional assumption $d_Q=\sqrt{n}$, this is possible only from the four parameters of Eqs.~\eqref{eq:c_minmax} and~\eqref{eq:a_minmax}, and hence provide a continuous method for certification.

Our main goal is to certify the presence of selected quantum measurements. Namely, notice that the inequality~\eqref{eq:dCboundDisc} can be saturated with a quantum system of dimension $d_Q=\sqrt{n}$ by considering a symmetric informationally complete quantum measurement (SIC-POVM), which is a measurement $M=\{\Pi_i/d_Q\}_{i=1}^{d_Q^2}$ composed of projectors $\Pi_i=\Pi_i^2$ such that $\tr(\Pi_i\Pi_j)=(d_Q\delta_{ij}+1)/(d_Q+1)$. Then $P_\alpha$ is obtained with $\alpha=1/d_Q$ by states $\varrho_j=\Pi_j$. This is the largest diagonal compatible with dimension $d_Q$, since $nc_{\min}\leq\tr(P)\leq d_Q$ by the information storability of $d_Q$-dimensional quantum systems~\cite{puliyil2025SDVIcomMats}. The question is whether SIC-POVMs are also necessary to saturate~\eqref{eq:dCboundDisc} with this diagonal. Indeed this is the case:
\begin{restatable}[SIC-class certification]{theorem}{selftestSIC}\label{thm:selftest}
Let $n=d_Q^2$. A $d_Q$-dimensional quantum strategy saturates the inequality~\eqref{eq:dCboundDisc} at $d_C=n$ with $c_{\min}=1/d_Q$ if and only if the states are pure, $\varrho_j=\dyad{\psi_j}$, and the measurements are $M_i=\dyad{\psi_i}/d_Q$ and form a SIC-POVM.
\end{restatable}

The proof is given in Appendix~\ref{app:SelfTest}. Notice that Theorem~\ref{thm:selftest} should be understood as a certification of SIC-class property, rather than as a self-test of a specified SIC representative. In the usual self-testing sense, one can only hope to identify a realization up to unavoidable gauge freedoms, such as a common unitary or antiunitary transformation and relabelling. Indeed, the statistics fix only the Hilbert-Schmidt Gram matrix of the projectors,
$\tr(\varrho_i\varrho_j)
 =|\braket{\psi_i}{\psi_j}|^2$,
and not the full complex Hilbert-space Gram matrix $\braket{\psi_i}{\psi_j}$. The information carried in the phases can be captured by higher-order invariants, such as the trace of products of three operators instead of two, and may distinguish inequivalent SIC-POVMs with identical pairwise transition probabilities~\cite{appleby2011lieSICunitConds,Tabia2013ExploringSICunitaryConds}.

For $d_Q=2$ the SIC is unique up to these symmetries, so that the measurement is identified; for $d_Q=3$ there is a continuous family of inequivalent SICs with identical pairwise overlaps~\cite{appleby2011lieSICunitConds,Tabia2013ExploringSICunitaryConds}, so that only the class is certified. This also shows that the uniqueness up to unitary or antiunitary equivalence stated in Proposition~3 of Ref.~\cite{puliyil2025SDVIcomMats} requires additional assumptions beyond maximal information storability, since for $d_Q=3$ inequivalent SICs yield the same communication matrix.

Notice that the anti-correlated inequality~\eqref{eq:dCboundExcl} can also be saturated at $d_C=n$ by using SIC-POVMs, with states $\varrho_j=(\one-\Pi_j)/(d_Q-1)$, which produce the optimal-exclusion CM $P_\alpha$ with $\alpha=0$. However, for $d_Q\geq3$ this does not lead to SIC certification, since the realization of $P_\alpha$ with $\alpha=0$ is not unique: the SIC states $\varrho_j=\Pi_j$ together with the rank-$(d_Q-1)$ effects $M_i=(\one-\Pi_i)/[d_Q(d_Q-1)]$ reproduce the same CM, and for odd $d_Q\geq5$ there are realizations in which neither the states nor the effects are rank one~\cite{LeeWeiDeWolf2017PSDrank}. Only for $d_Q=2$, where $\one-\Pi_i$ has rank one, is the exclusion realization unique: $\tr(\varrho_iM_i)=0$ forces both operators to be rank one, and the off-diagonal entries then fix the tetrahedron.

Theorem~\ref{thm:selftest} also offers an operational perspective on the open problem of SIC existence~\cite{appleby2011lieSICunitConds,Tabia2013ExploringSICunitaryConds}: a SIC-POVM exists in dimension $d_Q$ if and only if the statistics $P_\alpha$ with $\alpha=1/d_Q$ and $n=d_Q^2$ admit a $d_Q$-dimensional quantum realization. The existence problem can thus be investigated directly at the level of communication statistics, e.g., by numerically maximizing $\alpha$ over $d_Q$-dimensional strategies.

The minimal structure of Theorem~\ref{thm:selftest} also distinguishes it from previous dimension-bounded prepare-and-measure certification protocols. These typically use an input for Bob selecting among several measurement settings. For instance, \cite{Tavakoli2018SelfTestPM} develops robust self-tests of states and measurements, \cite{tavakoli2020self} focuses on robust self-testing of extremal qubit POVMs, \cite{Miklin2021universalSelfTestSIC} gives a general scheme certifying overlaps of pure states and projective measurements, with full robust self-testing for qubit configurations, and in \cite{Drotos2024SelfTestSICpovms}  a minimal prepare-and-measure self-test of a qubit semi-SIC POVM is provided. In contrast, the present construction uses only $d_Q^2$ preparations and a single $d_Q^2$-outcome measurement, and certifies SIC-class structure directly from the communication matrix.

Recent work on communication matrices and information storability provides a related structural perspective~\cite{puliyil2025SDVIcomMats}. Maximal information storability enforces rank-one state-effect alignment, whereas by saturating \eqref{eq:dCboundDisc} we further fix the statistics of SIC-POVMs. Moreover, this certification relies on a single inequality depending only on three extremal entries of $P$, rather than on its $d_Q^4$ individual probabilities. This provides a single figure of merit, whose saturation simultaneously certifies the SIC class and the quantum advantage $d_C=d_Q^2$ of the same statistics, and whose deviation from saturation directly controls the robustness of the certification, as we show next.

\section{Robustness for SIC property certification}{\label{sec:robustnes}}
In Theorem~\ref{thm:selftest} we assumed exact probability distributions. In order to address potential implementation, our next goal is to show that the certification tolerates small errors in the statistics and small leakage outside the assumed Hilbert space. In practical implementations, the probabilities $p(i|j)$ are not exact, but are affected by imperfect state preparations, noisy measurements, finite statistics, and possible leakage outside the intended Hilbert space. We first note that small implementation errors induce small perturbations of the observed communication matrix: errors $\Delta\varrho_j$ and $\Delta M_i$ in the states and measurements shift each probability by at most $\|\Delta\varrho_j\|_1\|M_i\|_\infty+\|\Delta M_i\|_\infty$ (Proposition~\ref{prop:continuity} in Appendix~\ref{app:robustness}). More importantly, near-extremal communication matrices imply approximate SIC structure.

\begin{restatable}[Robust SIC-POVM certification]{theorem}{robustSIC}\label{thm:robustSIC}
Let $P^{\rm SIC}$ denote the extremal SIC communication matrix of Theorem~\ref{thm:selftest}. Assume that a $d_Q$-dimensional realization $\{\varrho_i,M_i\}_{i=1}^{d_Q^2}$ generates a communication matrix $P$ satisfying $|P_{ij}-P^{\rm SIC}_{ij}| \le \varepsilon$ for all $i,j$. Then there exist rank-one projectors $\{\Pi_i\}_{i=1}^{d_Q^2}$ such that, for all $i,j$ and with $\kappa:=d_Q(d_Q^2+1)\varepsilon$,
\begin{gather}
    \|M_i-\Pi_i/d_Q\|_\infty\le d_Q^2\varepsilon,\qquad
    \|\varrho_i-\Pi_i\|_1\le 2\sqrt{\kappa},\nonumber\\
    \left|\tr(\Pi_i\Pi_j)-\frac{d_Q\delta_{ij}+1}{d_Q+1}\right|\le \kappa+2\sqrt{\kappa}.
\end{gather}
\end{restatable}
Hence every sufficiently near-extremal $d_Q$-dimensional realization is quantitatively close to a SIC-class realization. Notice that the assumption of Theorem~\ref{thm:robustSIC} only involves the four extremal entries: it is equivalent to $c_{\min}$ and $c_{\max}$ being $\varepsilon$-close to $1/d_Q$, and $a_{\min}$ and $a_{\max}$ being $\varepsilon$-close to $1/[d_Q(d_Q+1)]$. The proof is provided in Appendix~\ref{app:robustness}.

The exact dimension assumption in Theorem~\ref{thm:robustSIC} is essential, since additional uncontrolled degrees of freedom may support arbitrary measurement action without affecting the observed communication matrix. Nevertheless, robustness under small leakage can still be established: if the preparations leave the intended $d_Q$-dimensional subspace with probability at most $\delta$, the compressed realization satisfies the conclusions of Theorem~\ref{thm:robustSIC} with effective error parameter $\varepsilon+2\sqrt{\delta}+2\delta$ (Proposition~\ref{prop:leakage} in Appendix~\ref{app:robustness}). Therefore sufficiently small leakage preserves robust SIC-class certification.

\section{Dimension witness for quantum systems}
So far, the quantum dimension $d_Q$ was either assumed or upper bounded. Here we show that the inequalities of Theorem~\ref{thm:dCboundGeneral} also lower bound it from the four parameters used in this work alone, which provides a dimension witness in CS. Existing dimension witnesses rely on more structure than the setup of Fig.~\ref{fig:prepmeas}, such as several measurement settings, shared randomness or entanglement~\cite{Pal2008DimWitHighDimNonLoc,Brunner2008DimWItTestingnonloc,Vertesi2010DimWitLoopHoleNonLoc,Hendrych_2012DimWitnessExperiment}. In CS, the information storability of $d_Q$-dimensional quantum systems gives the linear bound $d_Q\geq\tr(P)$~\cite{puliyil2025SDVIcomMats}. However, this bound is only informative for correlated statistics: for anti-correlated CMs the trace is small and the bound becomes trivial. Moreover, as in the case of $d_C$, the sets of CMs with fixed $d_Q$ are not convex~\cite{gouveia2013psdRankConvex}, and therefore standard approaches to construct witnesses do not immediately apply.

The inequalities of Theorem~\ref{thm:dCboundGeneral} overcome this limitation as a natural consequence. Since $p(i|j)=\tr(\varrho_jM_i)$ is an inner product in the $d_Q^2$-dimensional real vector space of Hermitian operators, any quantum realization satisfies $\rank(P)\leq d_Q^2$~\cite{JdVicente2017SRandDVIdimWit}. Since the proof of Theorem~\ref{thm:dCboundGeneral} lower bounds $\rank(P)$ (Appendix~\ref{app:ProofGenIneq}), we naturally obtain the following dimension witness, a robust closed-form version of the bound $d_Q\geq\sqrt{\rank(P)}$.

\begin{observation}[Dimension witness]\label{obs:DimWit}
If an $n\times n$ communication matrix $P$ is obtained with quantum preparations and a quantum measurement acting on a Hilbert space of dimension $d_Q$, then
\begin{equation}\label{eq:DimWit}
    d_Q\geq\sqrt{1+\Gamma}\,.
\end{equation}
\end{observation}

Similarly as in Theorem~\ref{thm:dCboundGeneral}, this witness depends only on the extremal entries of $P$, and requires neither measurement incompatibility nor shared randomness. For the family $P_\alpha$ with $n=d_Q^2$ and $\alpha\neq 1/n$, the right-hand side of Eq.~\eqref{eq:DimWit} equals $\sqrt{n}$ (Lemma~\ref{lem:maxSaturationPalpha}). The bound is thus tight at both limitations imposed by quantum mechanics: optimal discrimination ($\alpha=1/\sqrt{n}$) is attained by a SIC-POVM with states $\varrho_j=\Pi_j$, and optimal exclusion ($\alpha=0$) by the same SIC-POVM with states $\varrho_j=(\one-\Pi_j)/(d_Q-1)$, both in dimension $d_Q=\sqrt{n}$. Noise robustness follows similarly as in the results above:
if each entry of the experimental CM is known up to an error $\varepsilon$, evaluating Eq.~\eqref{eq:DimWit} with $c_{\min}-\varepsilon$, $a_{\min}-\varepsilon$ and $c_{\max}+\varepsilon$ yields a valid dimension witness.

\section{Conclusions}
We have introduced inequalities for the classical and quantum dimensions needed to reproduce communication statistics. These results characterize regions of CMs exhibiting a gap between the classical and quantum dimension, rather than isolated examples. In turn, they demonstrate that two operational strategies display such a gap: correlations (given by state discrimination), and anticorrelations (given by state exclusion). We have shown that saturation of one of the inequalities occurs only when SIC-POVMs are used, and therefore provided an explicit protocol to certify these measurements with communication matrices under dimensional assumptions. We have also shown that this certification is robust to noise and leakage.

While existing literature has mostly focused on anti-correlated messages to display quantum advantage, our work shows that correlated messages display the same quantum advantage and are better suited for certification.
Despite non-convexity of the problem, the presented approach provides robust witnesses for the Hilbert space dimension of the underlying quantum system, valid for both correlated and anti-correlated statistics.
Robust quantum certification is thus possible in this minimalistic setting, and the results obtained are specifically tailored for a realistic implementation and scalable near-term quantum communication scenarios, where practical constraints demand a simplified structure and minimal use of shared and local resources.

\bigskip

{\em Acknowledgements.\,\,} We are thankful to Arnau Diebra, Carlos de Gois,  Matthias Kleinmann, Leevi Lepp\"aj\"arvi, Ramon Muñoz-Tapia, Marc Olivier Renou, Jef Pauwels, Davide Poderini, Alejandro Pozas-Kerstjens, Anna Sanpera and Lucas Tendick for discussions. AR acknowledges financial support from the Deutsche Forschungsgemeinschaft (DFG, German
Research Foundation, project number 563437167).
JC acknowledges support from ICREA Academia award.
AR, MZ and JC acknowledge financial support from Spanish MICIN (projects PID2022-141283NB-I00 and PID2022-139099NB-I00) with the support of FEDER funds, the Spanish Government with funding from European Union NextGenerationEU (PRTR-C17.I1), the Generalitat de Catalunya, the Ministry for Digital Transformation and of Civil Service of the Spanish Government through the QUANTUM ENIA project -Quantum Spain Project- through the Recovery, Transformation and Resilience Plan NextGeneration EU within the framework of the Digital Spain 2026 Agenda. JF acknowledges funding from MICIU/AEI/10.13039/501100011033 under Grant PID2024-160393NB-I00. MZ also acknowledges
support from Ministerio de Ciencia e Innovación of the Spanish Government PRE2020-093634. SSB acknowledges financial support from the Beatriu de Pinós programme of the Ministry of Research and Universities of the Government of Catalonia (grant number 2024 BP 00267).

\bigskip

\bigskip

{\em AI and human statement.} Gemini was used in August 2025 to identify ergodicity coefficients and their relation to eigenvalues. This closed a gap in the proof of Eq.~\eqref{eq:dCboundExcl}, which was incomplete until then.
In October 2026, Claude Opus 5.5 was used to prove Proposition~\ref{prop:dQupper}, which we did not manage to prove earlier without the assumption of existence of SIC-POVMs. The proof is mostly done by AI, approved and edited by us, and replaced an earlier weaker result. Theorem~\ref{thm:robustSIC} was derived hybridly by interaction with Claude. Figs.~\ref{fig:Sets-dQ-dC} and~\ref{fig:Palpha} are assisted with Claude. The scope, ideas, proof strategies, and writing are human.

\addcontentsline{toc}{subsection}{Bibliography}
\bibliographystyle{apsrev4-2}
\bibliography{Bibliography}{}

@article{brunner2014bell,
  title={Bell nonlocality},
  author={Brunner, Nicolas and Cavalcanti, Daniel and Pironio, Stefano and Scarani, Valerio and Wehner, Stephanie},
  journal={Rev. Mod. Phys.},
  volume={86},
  number={2},
  pages={419--478},
  year={2014},
  publisher={APS},
  doi={10.1103/RevModPhys.86.419}
}

@article{mattle1996dense,
  title={Dense coding in experimental quantum communication},
  author={Mattle, Klaus and Weinfurter, Harald and Kwiat, Paul G and Zeilinger, Anton},
  journal={Phys. Rev. Lett.},
  volume={76},
  number={25},
  pages={4656},
  year={1996},
  publisher={APS},
  doi={10.1103/PhysRevLett.76.4656}
}

@article{Ding2024quantum,
  title={Quantum advantage: a single qubit’s experimental edge in classical data storage},
  author={Ding, Chen and Lobo, Edwin Peter and Alimuddin, Mir and Xu, Xiao-Yue and Zhang, Shuo and Banik, Manik and Bao, Wan-Su and Huang, He-Liang},
  journal={Phys. Rev. Lett.},
  volume={133},
  number={20},
  pages={200201},
  year={2024},
  publisher={APS},
  doi={10.1103/PhysRevLett.133.200201}
}

@article{Seneta1948Explicit,
title = {Explicit forms for ergodicity coefficients and spectrum localization},
journal = {Linear Algebra and its Applications},
volume = {60},
pages = {187-197},
year = {1984},
issn = {0024-3795},
doi = {https://doi.org/10.1016/0024-3795(84)90079-X},
url = {https://www.sciencedirect.com/science/article/pii/002437958490079X},
author = {E. Seneta}
}

@article{Ipsen2011ErCoeff,
author = {Ipsen, Ilse and Selee, Teresa},
year = {2011},
month = {01},
pages = {153-200},
title = {Ergodicity Coefficients Defined by Vector Norms},
volume = {32},
journal = {SIAM J. Matrix Analysis Applications},
doi = {10.1137/090752948}
}

@article{Heino24SimpleUnboundQAdvtg,
  title = {Simple information-processing tasks with unbounded quantum advantage},
  author = {Heinosaari, Teiko and Kerppo, Oskari and Lepp\"aj\"arvi, Leevi and Pl\'avala, Martin},
  journal = {Phys. Rev. A},
  volume = {109},
  issue = {3},
  pages = {032627},
  numpages = {8},
  year = {2024},
  month = {Mar},
  publisher = {American Physical Society},
  doi = {10.1103/PhysRevA.109.032627},
  url = {https://link.aps.org/doi/10.1103/PhysRevA.109.032627}
}

@article{RandFree_23_Some,
  title = {Randomness-Free Test of Nonclassicality: A Proof of Concept},
  author = {Ma, Zhonghua and Rambach, Markus and Goswami, Kaumudibikash and Bhattacharya, Some Sankar and Banik, Manik and Romero, Jacquiline},
  journal = {Phys. Rev. Lett.},
  volume = {131},
  issue = {13},
  pages = {130201},
  numpages = {7},
  year = {2023},
  month = {Sep},
  publisher = {American Physical Society},
  doi = {10.1103/PhysRevLett.131.130201},
  url = {https://link.aps.org/doi/10.1103/PhysRevLett.131.130201}
}

@article{Shor1995PolynomialTimeAF,
  title={Polynomial-Time Algorithms for Prime Factorization and Discrete Logarithms on a Quantum Computer},
  author={Peter W. Shor},
  journal={SIAM Rev.},
  year={1999},
  volume={41},
  pages={303-332},
  url={https://api.semanticscholar.org/CorpusID:2337707}
}

@inproceedings{Grover1996Search,
author = {Grover, Lov K.},
title = {A fast quantum mechanical algorithm for database search},
year = {1996},
isbn = {0897917855},
publisher = {Association for Computing Machinery},
address = {New York, NY, USA},
url = {https://doi.org/10.1145/237814.237866},
doi = {10.1145/237814.237866},
booktitle = {Proceedings of the Twenty-Eighth Annual ACM Symposium on Theory of Computing},
pages = {212–219},
numpages = {8},
location = {Philadelphia, Pennsylvania, USA},
series = {STOC '96}
}

@article{zhou2018HLimitNature,
  title={Achieving the Heisenberg limit in quantum metrology using quantum error correction},
  author={Zhou, Sisi and Zhang, Mengzhen and Preskill, John and Jiang, Liang},
  journal={Nat. Commun.},
  volume={9},
  number={1},
  pages={78},
  year={2018},
  publisher={Nature Publishing Group UK London},
  doi={10.1038/s41467-017-02510-3}
}

@misc{puliyil2025SDVIcomMats,
      title={Semi-device-independent channel identification with communication matrices}, 
      author={Samgeeth Puliyil and Leevi Leppäjärvi and Mário Ziman},
      year={2025},
      eprint={2511.14273},
      archivePrefix={arXiv},
      primaryClass={quant-ph},
      url={https://arxiv.org/abs/2511.14273}, 
}

@article{Supic2020Self-testReview,
  doi = {10.22331/q-2020-09-30-337},
  url = {https://doi.org/10.22331/q-2020-09-30-337},
  title = {Self-testing of quantum systems: a review},
  author = {{\v{S}}upi{\'{c}}, Ivan and Bowles, Joseph},
  journal = {{Quantum}},
  issn = {2521-327X},
  publisher = {{Verein zur F{\"{o}}rderung des Open Access Publizierens in den Quantenwissenschaften}},
  volume = {4},
  pages = {337},
  month = sep,
  year = {2020}
}

@article{Heinosaari2024maxElsCommuns,
  doi = {10.22331/q-2024-11-07-1515},
  url = {https://doi.org/10.22331/q-2024-11-07-1515},
  title = {Maximal {E}lements of {Q}uantum {C}ommunication},
  author = {Heinosaari, Teiko and Kerppo, Oskari},
  journal = {{Quantum}},
  issn = {2521-327X},
  publisher = {{Verein zur F{\"{o}}rderung des Open Access Publizierens in den Quantenwissenschaften}},
  volume = {8},
  pages = {1515},
  month = nov,
  year = {2024}
}

@article{jain2013efficient,
  title={Efficient protocols for generating bipartite classical distributions and quantum states},
  author={Jain, Rahul and Shi, Yaoyun and Wei, Zhaohui and Zhang, Shengyu},
  journal={IEEE Transactions on Information Theory},
  volume={59},
  number={8},
  pages={5171--5178},
  year={2013},
  publisher={IEEE},
  doi={10.1109/TIT.2013.2258372}
}

@article{Fawzi_2014LBnnegRank,
   title={Lower bounds on nonnegative rank via nonnegative nuclear norms},
   volume={153},
   ISSN={1436-4646},
   url={http://dx.doi.org/10.1007/s10107-014-0837-2},
   DOI={10.1007/s10107-014-0837-2},
   number={1},
   journal={Mathematical Programming},
   publisher={Springer Science and Business Media LLC},
   author={Fawzi, Hamza and Parrilo, Pablo A.},
   year={2014},
   month=nov, pages={41–66} }

@article{Uola2020ResourceExclusion,
  title = {All Quantum Resources Provide an Advantage in Exclusion Tasks},
  author = {Uola, Roope and Bullock, Tom and Kraft, Tristan and Pellonp\"a\"a, Juha-Pekka and Brunner, Nicolas},
  journal = {Phys. Rev. Lett.},
  volume = {125},
  issue = {11},
  pages = {110402},
  numpages = {6},
  year = {2020},
  month = {Sep},
  publisher = {American Physical Society},
  doi = {10.1103/PhysRevLett.125.110402},
  url = {https://link.aps.org/doi/10.1103/PhysRevLett.125.110402}
}

@article{Drotos2024SelfTestSICpovms,
  title = {Self-testing of semisymmetric informationally complete measurements in a qubit prepare-and-measure scenario},
  author = {Dr\'otos, G\'abor and P\'al, K\'aroly F. and V\'ertesi, Tam\'as},
  journal = {Phys. Rev. A},
  volume = {110},
  issue = {3},
  pages = {032427},
  numpages = {11},
  year = {2024},
  month = {Sep},
  publisher = {American Physical Society},
  doi = {10.1103/PhysRevA.110.032427},
  url = {https://link.aps.org/doi/10.1103/PhysRevA.110.032427}
}

@article{Miklin2021universalSelfTestSIC,
  doi = {10.22331/q-2021-04-06-424},
  url = {https://doi.org/10.22331/q-2021-04-06-424},
  title = {A universal scheme for robust self-testing in the prepare-and-measure scenario},
  author = {Miklin, Nikolai and Oszmaniec, Micha{\l{}}},
  journal = {{Quantum}},
  issn = {2521-327X},
  publisher = {{Verein zur F{\"{o}}rderung des Open Access Publizierens in den Quantenwissenschaften}},
  volume = {5},
  pages = {424},
  month = apr,
  year = {2021}
}

@article{Guhne_2009EntDet,
   title={Entanglement detection},
   volume={474},
   ISSN={0370-1573},
   url={http://dx.doi.org/10.1016/j.physrep.2009.02.004},
   DOI={10.1016/j.physrep.2009.02.004},
   number={1–6},
   journal={Physics Reports},
   publisher={Elsevier BV},
   author={Gühne, Otfried and Tóth, Géza},
   year={2009},
   month=apr, pages={1–75} }

@article{Contextuality2022Review,
  title = {Kochen-Specker contextuality},
  author = {Budroni, Costantino and Cabello, Ad\'an and G\"uhne, Otfried and Kleinmann, Matthias and Larsson, Jan-\AA{}ke},
  journal = {Rev. Mod. Phys.},
  volume = {94},
  issue = {4},
  pages = {045007},
  numpages = {62},
  year = {2022},
  month = {Dec},
  publisher = {American Physical Society},
  doi = {10.1103/RevModPhys.94.045007},
  url = {https://link.aps.org/doi/10.1103/RevModPhys.94.045007}
}

@article{Emary_2014Legggett-GargIneq,
doi = {10.1088/0034-4885/77/1/016001},
url = {https://doi.org/10.1088/0034-4885/77/1/016001},
journal = {Rep. Prog. Phys.},
year = {2014},
month = {dec},
publisher = {IOP Publishing},
volume = {77},
number = {1},
pages = {016001},
author = {Emary, Clive and Lambert, Neill and Nori, Franco},
title = {Leggett–Garg inequalities}
}

@article{Fritz_2010TemporalCHSH,
doi = {10.1088/1367-2630/12/8/083055},
url = {https://doi.org/10.1088/1367-2630/12/8/083055},
year = {2010},
month = {aug},
publisher = {},
volume = {12},
number = {8},
pages = {083055},
author = {Fritz, Tobias},
title = {Quantum correlations in the temporal Clauser–Horne–Shimony–Holt (CHSH) scenario},
journal = {New Journal of Physics}
}

@article{Egelhaaf_2025CertIncPrepMeasBell,
doi = {10.1088/1751-8121/adb3ff},
url = {https://doi.org/10.1088/1751-8121/adb3ff},
year = {2025},
month = {mar},
publisher = {IOP Publishing},
volume = {58},
number = {9},
pages = {095304},
author = {Egelhaaf, Sophie and Pauwels, Jef and Túlio Quintino, Marco and Uola, Roope},
title = {Certifying measurement incompatibility in prepare-and-measure and Bell scenarios},
journal = {Journal of Physics A: Mathematical and Theoretical}
}

@article{Tavakoli2018SelfTestPM,
  title = {Self-testing quantum states and measurements in the prepare-and-measure scenario},
  author = {Tavakoli, Armin and Kaniewski, J{\k{e}}drzej and V\'ertesi, Tam\'as and Rosset, Denis and Brunner, Nicolas},
  journal = {Phys. Rev. A},
  volume = {98},
  issue = {6},
  pages = {062307},
  numpages = {13},
  year = {2018},
  month = {Dec},
  publisher = {American Physical Society},
  doi = {10.1103/PhysRevA.98.062307},
  url = {https://link.aps.org/doi/10.1103/PhysRevA.98.062307}
}

@article{tavakoli2020self,
  title={Self-testing nonprojective quantum measurements in prepare-and-measure experiments},
  author={Tavakoli, Armin and Smania, Massimiliano and V{\'e}rtesi, Tam{\'a}s and Brunner, Nicolas and Bourennane, Mohamed},
  journal={Science Advances},
  volume={6},
  number={16},
  pages={eaaw6664},
  year={2020},
  publisher={American Association for the Advancement of Science},
  url = {https://www.science.org/doi/full/10.1126/sciadv.aaw6664}
}

@article{Chakraborty2025AdvQubitComPM,
doi = {10.1088/1367-2630/adb174},
url = {https://doi.org/10.1088/1367-2630/adb174},
year = {2025},
month = {feb},
publisher = {IOP Publishing},
volume = {27},
number = {2},
pages = {023027},
author = {Chakraborty, Ananya and Gopalkrishna Naik, Sahil and Lobo, Edwin Peter and Patra, Ram Krishna and Sen, Samrat and Alimuddin, Mir and Mukherjee, Amit and Banik, Manik},
title = {Unlocking the advantage of qubit communication in multi-node network configurations},
journal = {New Journal of Physics}
}

@misc{doolittle2025OpFrQComNettsPM,
      title={An Operational Framework for Nonclassicality in Quantum Communication Networks}, 
      author={Brian Doolittle and Felix Leditzky and Eric Chitambar},
      year={2025},
      eprint={2403.02988},
      archivePrefix={arXiv},
      primaryClass={quant-ph},
      url={https://arxiv.org/abs/2403.02988}, 
}

@article{Frenkel_2015ComMats,
   title={Classical Information Storage in an n-Level Quantum System},
   volume={340},
   ISSN={1432-0916},
   url={http://dx.doi.org/10.1007/s00220-015-2463-0},
   DOI={10.1007/s00220-015-2463-0},
   number={2},
   journal={Communications in Mathematical Physics},
   publisher={Springer Science and Business Media LLC},
   author={Frenkel, Péter E. and Weiner, Mihály},
   year={2015},
   month=sep, pages={563–574} }

@article{Heinosaari_2020ComMats,
   title={Communication tasks in operational theories},
   volume={53},
   ISSN={1751-8121},
   url={http://dx.doi.org/10.1088/1751-8121/abb5dc},
   DOI={10.1088/1751-8121/abb5dc},
   number={43},
   journal={Journal of Physics A: Mathematical and Theoretical},
   publisher={IOP Publishing},
   author={Heinosaari, Teiko and Kerppo, Oskari and Leppäjärvi, Leevi},
   year={2020},
   month=oct, pages={435302} }

@article{cohen1993NNrank,
  title={Nonnegative ranks, decompositions, and factorizations of nonnegative matrices},
  author={Cohen, Joel E and Rothblum, Uriel G},
  journal={Linear Algebra and its Applications},
  volume={190},
  pages={149--168},
  year={1993},
  publisher={Elsevier},
url={https://www.sciencedirect.com/science/article/pii/002437959390224C}
}

@article{shitov2017complexityNNRank,
  title={The complexity of positive semidefinite matrix factorization},
  author={Shitov, Yaroslav},
  journal={SIAM Journal on Optimization},
  volume={27},
  number={3},
  pages={1898--1909},
  year={2017},
  publisher={SIAM},
  url={https://epubs.siam.org/doi/abs/10.1137/16M1080616}
}

@article{vavasis2010complexityNNrank,
  title={On the complexity of nonnegative matrix factorization},
  author={Vavasis, Stephen A},
  journal={SIAM journal on optimization},
  volume={20},
  number={3},
  pages={1364--1377},
  year={2010},
  publisher={SIAM},
  url={https://epubs.siam.org/doi/abs/10.1137/070709967}
}

@article{gillis2012geometricNNrank,
  title={On the geometric interpretation of the nonnegative rank},
  author={Gillis, Nicolas and Glineur, Fran{\c{c}}ois},
  journal={Linear Algebra and its Applications},
  volume={437},
  number={11},
  pages={2685--2712},
  year={2012},
  publisher={Elsevier},
  url={https://www.sciencedirect.com/science/article/pii/S0024379512005022}
}

@article{shitov2014upperBoundNNrank,
  title={An upper bound for nonnegative rank},
  author={Shitov, Yaroslav},
  journal={Journal of Combinatorial Theory, Series A},
  volume={122},
  pages={126--132},
  year={2014},
  publisher={Elsevier},
  url={https://www.sciencedirect.com/science/article/pii/S0097316513001635}
}

@article{Brun2002QSExcl,
  title = {How much state assignments can differ},
  author = {Brun, Todd A. and Finkelstein, J. and Mermin, N. David},
  journal = {Phys. Rev. A},
  volume = {65},
  issue = {3},
  pages = {032315},
  numpages = {4},
  year = {2002},
  month = {Feb},
  publisher = {American Physical Society},
  doi = {10.1103/PhysRevA.65.032315},
  url = {https://link.aps.org/doi/10.1103/PhysRevA.65.032315}
}

@inproceedings{fiorini2012NNrankdC,
  title={Linear vs. semidefinite extended formulations: exponential separation and strong lower bounds},
  author={Fiorini, Samuel and Massar, Serge and Pokutta, Sebastian and Tiwary, Hans Raj and De Wolf, Ronald},
  booktitle={Proceedings of the forty-fourth annual ACM symposium on Theory of computing},
  pages={95--106},
  year={2012},
  url={https://dl.acm.org/doi/abs/10.1145/2213977.2213988}
}

@article{Hendrych_2012DimWitnessExperiment,
   title={Experimental estimation of the dimension of classical and quantum systems},
   volume={8},
   ISSN={1745-2481},
   url={http://dx.doi.org/10.1038/nphys2334},
   DOI={10.1038/nphys2334},
   number={8},
   journal={Nature Physics},
   publisher={Springer Science and Business Media LLC},
   author={Hendrych, Martin and Gallego, Rodrigo and Mičuda, Michal and Brunner, Nicolas and Acín, Antonio and Torres, Juan P.},
   year={2012},
   month=jun, pages={588–591} }

@article{Brunner2008DimWItTestingnonloc,
  title = {Testing the Dimension of Hilbert Spaces},
  author = {Brunner, Nicolas and Pironio, Stefano and Acin, Antonio and Gisin, Nicolas and M\'ethot, Andr\'e Allan and Scarani, Valerio},
  journal = {Phys. Rev. Lett.},
  volume = {100},
  issue = {21},
  pages = {210503},
  numpages = {4},
  year = {2008},
  month = {May},
  publisher = {American Physical Society},
  doi = {10.1103/PhysRevLett.100.210503},
  url = {https://link.aps.org/doi/10.1103/PhysRevLett.100.210503}
}

@article{Pal2008DimWitHighDimNonLoc,
  title = {Efficiency of higher-dimensional Hilbert spaces for the violation of Bell inequalities},
  author = {P\'al, K\'aroly F. and V\'ertesi, Tam\'as},
  journal = {Phys. Rev. A},
  volume = {77},
  issue = {4},
  pages = {042105},
  numpages = {9},
  year = {2008},
  month = {Apr},
  publisher = {American Physical Society},
  doi = {10.1103/PhysRevA.77.042105},
  url = {https://link.aps.org/doi/10.1103/PhysRevA.77.042105}
}

@article{Vertesi2010DimWitLoopHoleNonLoc,
  title = {Closing the Detection Loophole in Bell Experiments Using Qudits},
  author = {V\'ertesi, Tam\'as and Pironio, Stefano and Brunner, Nicolas},
  journal = {Phys. Rev. Lett.},
  volume = {104},
  issue = {6},
  pages = {060401},
  numpages = {4},
  year = {2010},
  month = {Feb},
  publisher = {American Physical Society},
  doi = {10.1103/PhysRevLett.104.060401},
  url = {https://link.aps.org/doi/10.1103/PhysRevLett.104.060401}
}

@article{holevo1973bounds,
  title={Bounds for the quantity of information transmitted by a quantum communication channel},
  author={Holevo, Alexander Semenovich},
  journal={Problemy Peredachi Informatsii},
  volume={9},
  number={3},
  pages={3--11},
  year={1973},
  publisher={Russian Academy of Sciences, Branch of Informatics, and Computer Equipment and},
  url={https://www.mathnet.ru/eng/ppi903}
}

@article{Barnett2009QSDisc,
author = {Stephen M. Barnett and Sarah Croke},
journal = {Adv. Opt. Photon.},
number = {2},
pages = {238--278},
publisher = {Optica Publishing Group},
title = {Quantum state discrimination},
volume = {1},
month = {Apr},
year = {2009},
url = {https://opg.optica.org/aop/abstract.cfm?URI=aop-1-2-238},
doi = {10.1364/AOP.1.000238}
}

@article{Bae2015ASDiscApps,
   title={Quantum state discrimination and its applications},
   volume={48},
   ISSN={1751-8121},
   url={http://dx.doi.org/10.1088/1751-8113/48/8/083001},
   DOI={10.1088/1751-8113/48/8/083001},
   number={8},
   journal={Journal of Physics A: Mathematical and Theoretical},
   publisher={IOP Publishing},
   author={Bae, Joonwoo and Kwek, Leong-Chuan},
   year={2015},
   month=jan, pages={083001} }

@article{Bandyopadhyay2014QSExcl,
   title={Conclusive exclusion of quantum states},
   volume={89},
   ISSN={1094-1622},
   url={http://dx.doi.org/10.1103/PhysRevA.89.022336},
   DOI={10.1103/physreva.89.022336},
   number={2},
   journal={Phys. Rev. A},
   pages={022336},
   publisher={American Physical Society (APS)},
   author={Bandyopadhyay, Somshubhro and Jain, Rahul and Oppenheim, Jonathan and Perry, Christopher},
   year={2014},
   month=feb }

@inproceedings{coppersmith1987RankPoly2,
  title={Matrix multiplication via arithmetic progressions},
  author={Coppersmith, Don and Winograd, Shmuel},
  booktitle={Proceedings of the nineteenth annual ACM symposium on Theory of computing},
  pages={1--6},
  year={1987},
  doi={10.1145/28395.28396}
}

@article{Alman_2024RankPoly2,
   title={A Refined Laser Method and Faster Matrix Multiplication},
   volume={3},
   pages={21},
   ISSN={2751-4838},
   url={http://dx.doi.org/10.46298/theoretics.24.21},
   DOI={10.46298/theoretics.24.21},
   journal={TheoretiCS},
   publisher={Centre pour la Communication Scientifique Directe (CCSD)},
   author={Alman, Josh and Williams, Virginia Vassilevska},
   year={2024},
   month=sep }

@misc{ExPaper,
      title={Robust certification of high-dimensional quantum devices}, 
      author={Javier Fernández and Albert Rico and David Viedma and Evelyn A. Ortega and Valerio Pruneri and Adam Vallés and Verònica Ahufinger and Anna Sanpera and Some S. Bhattacharya},
      year={2026},
      eprint={2605.04338},
      archivePrefix={arXiv},
      primaryClass={quant-ph},
      url={https://arxiv.org/abs/2605.04338}, 
}

@article{VanHimbeeck2017SDVIPMwithRandomness,
  doi = {10.22331/q-2017-11-18-33},
  url = {https://doi.org/10.22331/q-2017-11-18-33},
  title = {Semi-device-independent framework based on natural physical assumptions},
  author = {Van Himbeeck, Thomas and Woodhead, Erik and Cerf, Nicolas J. and Garc{\'{i}}a-Patr{\'{o}}n, Ra{\'{u}}l and Pironio, Stefano},
  journal = {{Quantum}},
  issn = {2521-327X},
  publisher = {{Verein zur F{\"{o}}rderung des Open Access Publizierens in den Quantenwissenschaften}},
  volume = {1},
  pages = {33},
  month = nov,
  year = {2017}
}

@article{Designolle2021CoherenceSet,
  title = {Set Coherence: Basis-Independent Quantification of Quantum Coherence},
  author = {Designolle, S\'ebastien and Uola, Roope and Luoma, Kimmo and Brunner, Nicolas},
  journal = {Phys. Rev. Lett.},
  volume = {126},
  issue = {22},
  pages = {220404},
  numpages = {6},
  year = {2021},
  month = {Jun},
  publisher = {American Physical Society},
  doi = {10.1103/PhysRevLett.126.220404},
  url = {https://link.aps.org/doi/10.1103/PhysRevLett.126.220404}
}

@article{Gois2021IneqPM,
  title = {General Method for Classicality Certification in the Prepare and Measure Scenario},
  author = {de Gois, Carlos and Moreno, George and Nery, Ranieri and Brito, Samura\'{\i} and Chaves, Rafael and Rabelo, Rafael},
  journal = {PRX Quantum},
  volume = {2},
  issue = {3},
  pages = {030311},
  numpages = {12},
  year = {2021},
  month = {Jul},
  publisher = {American Physical Society},
  doi = {10.1103/PRXQuantum.2.030311},
  url = {https://link.aps.org/doi/10.1103/PRXQuantum.2.030311}
}

@misc{brask2026IneqPM,
      title={Quantum correlations in prepare-and-measure scenarios and their semi-device-independent applications}, 
      author={Jonatan Bohr Brask and Nicolas Brunner and Jef Pauwels and Davide Rusca and Armin Tavakoli},
      year={2026},
      eprint={2603.23604},
      archivePrefix={arXiv},
      primaryClass={quant-ph},
      url={https://arxiv.org/abs/2603.23604}, 
}

@article{JdVicente2017SRandDVIdimWit,
  title = {Shared randomness and device-independent dimension witnessing},
  author = {de Vicente, Julio I.},
  journal = {Phys. Rev. A},
  volume = {95},
  issue = {1},
  pages = {012340},
  numpages = {11},
  year = {2017},
  month = {Jan},
  publisher = {American Physical Society},
  doi = {10.1103/PhysRevA.95.012340},
  url = {https://link.aps.org/doi/10.1103/PhysRevA.95.012340}
}

@article{gouveia2013psdRankConvex,
  title={Polytopes of minimum positive semidefinite rank},
  author={Gouveia, Joao and Robinson, Richard Z and Thomas, Rekha R},
  journal={Discrete \& Computational Geometry},
  volume={50},
  number={3},
  pages={679--699},
  year={2013},
  publisher={Springer},
  doi={10.1007/s00454-013-9533-x}
}

@article{fawzi2015psdRank,
  title={Positive semidefinite rank},
  author={Fawzi, Hamza and Gouveia, Jo{\~a}o and Parrilo, Pablo A and Robinson, Richard Z and Thomas, Rekha R},
  journal={Mathematical Programming},
  volume={153},
  number={1},
  pages={133--177},
  year={2015},
  publisher={Springer},
  doi={10.1007/s10107-015-0922-1}
}

@article{appleby2011lieSICunitConds,
  title={The Lie algebraic significance of symmetric informationally complete measurements},
  author={Appleby, DM and Flammia, Steven T and Fuchs, Christopher A},
  journal={J. Math. Phys.},
  volume={52},
  number={2},
  pages={022202},
  year={2011},
  publisher={AIP Publishing},
  doi={10.1063/1.3555805}
}

@article{Tabia2013ExploringSICunitaryConds,
  title = {Exploring the geometry of qutrit state space using symmetric informationally complete probabilities},
  author = {Tabia, Gelo Noel M. and Appleby, D. M.},
  journal = {Phys. Rev. A},
  volume = {88},
  issue = {1},
  pages = {012131},
  numpages = {8},
  year = {2013},
  month = {Jul},
  publisher = {American Physical Society},
  doi = {10.1103/PhysRevA.88.012131},
  url = {https://link.aps.org/doi/10.1103/PhysRevA.88.012131}
}

@article{Hahn1927,
  author  = {Hahn, Hans},
  title   = {{\"U}ber lineare {G}leichungssysteme in linearen {R}{\"a}umen},
  journal = {Journal f{\"u}r die reine und angewandte Mathematik},
  volume  = {157},
  pages   = {214--229},
  year    = {1927},
  doi     = {10.1515/crll.1927.157.214}
}

@article{Banach1929,
  author  = {Banach, Stefan},
  title   = {Sur les fonctionnelles lin{\'e}aires},
  journal = {Studia Mathematica},
  volume  = {1},
  pages   = {211--216},
  year    = {1929},
  doi={10.4064/sm-1-1-211-216}
}

@book{HornJohnson2012,
  author = {Horn, Roger A. and Johnson, Charles R.},
  title = {Matrix Analysis}, edition = {2},
  publisher = {Cambridge University Press},
  year = {2012},
  doi={10.1017/CBO9781139020411}
}

@article{LeeWeiDeWolf2017PSDrank,
  title = {Some upper and lower bounds on {PSD}-rank},
  author = {Lee, Troy and Wei, Zhaohui and de Wolf, Ronald},
  journal = {Mathematical Programming},
  volume = {162},
  pages = {495--521},
  year = {2017},
  eprint = {1407.4308},
  archivePrefix = {arXiv}
}

\begin{widetext}
\appendix

\section{Proof of Theorem~\ref{thm:dCboundGeneral}}\label{app:ProofGenIneq}
\dCboundGeneral*

The dimension $d_C$ needed to simulate $P$ with classical labels is known to be equal to the non-negative rank of $P$~\cite{Heino24SimpleUnboundQAdvtg}. Although this parameter is difficult to compute, it is lower bounded by the rank $r$ of $P$. Therefore it is enough to lower bound the rank. Now we proceed to the proof that the rank of $P$ is lower bounded as in Theorem~\ref{thm:dCboundGeneral}.
\begin{proof}

Let $P$ be stochastic, and let $\lambda_1,...,\lambda_n$ be its (complex) eigenvalues, sorted in non-increasing order of their absolute value. It follows from the Perron-Frobenius theorem that $\lambda_1=1$ (since the all-ones vector is a left eigenvector of any column-stochastic matrix) and $|\lambda_{i\neq 1}|\leq 1$. At the same time, $\tr(P)\geq 1-|\lambda_2|(r-1)$, where $r$ is the rank of $P$, saturated by the case where all eigenvalues are real, negative and as large as the second eigenvalue $\lambda_2$ in absolute value. In turn, this means that $r\geq (1-\tr(P))/|\lambda_2|+1$. Therefore we expect the rank to be large for anti-correlated CMs (which have low trace) with small $|\lambda_2|$. Our goal is then to find an upper bound for $|\lambda_2|$ from the entries of $P$, which will in turn lower bound the number of nonzero eigenvalues $r$.

For that, we will use the following relation in~\cite{Seneta1948Explicit,Ipsen2011ErCoeff}. Let $S=\{x\in\mathbf{R}_+^n:\,\sum_ix_i=1\}$ and define
\begin{equation}
    \tau_p(P) = \sup_{x,y\in S:\,x\neq y} \frac{||Px-Py||_p}{||x-y||_p}\,,
\end{equation}
which are called {\em ergodicity coefficients}~\cite{Seneta1948Explicit}. The assumption that $P$ is stochastic implies that the coefficient $\tau_1$ is given by~\cite{Seneta1948Explicit}
\begin{equation}\label{eq:Tau1Expr}
    \tau_1(P)=1-\min_{j\neq k}\sum_i\min\{P_{ij},P_{ik}\}
\end{equation}
and satisfies $\tau_1(P)\geq |\lambda_i|$ for all eigenvalues $\lambda_i$ of $P$ with $i\neq 1$~\cite{Ipsen2011ErCoeff}.

Now our goal is to upper bound $\tau_1(P)$ from the entries of $P$. Let us focus on each individual term of the right hand side in Eq.~\eqref{eq:Tau1Expr}. For each pair of distinct columns $j\neq k$, we compare $P_{ij}$ and $P_{ik}$ row by row. We have three cases:
When $i=j$, we have $\min\{P_{jj},P_{jk}\}\geq\min\{c_{\min},a_{\min}\}$, since $P_{jj}\geq c_{\min}$ and $P_{jk}\geq a_{\min}$.
When $i=k$, similarly $\min\{P_{kj},P_{kk}\}\geq\min\{a_{\min},c_{\min}\}$.
And for $j\neq i\neq k$, we are comparing two off-diagonal terms and therefore $\min\{P_{ij},P_{ik}\}\geq a_{\min}$.
The first two cases will occur two times over the whole summation, and the last case will occur $n-2$ times. Therefore we have
\begin{equation}\label{eq:1-tau}
    1-\tau_1(P)=\min_{j\neq k}\sum_i\min\{P_{ij},P_{ik}\} \geq
    (n-2)a_{\min} + 2\min\{a_{\min},c_{\min}\}
\end{equation}
In turn this means that $\tau_1(P)\leq 1-(n-2)a_{\min} - 2\min\{a_{\min},c_{\min}\}$, which, since $|\lambda_2|\leq\tau_1(P)$, implies
\begin{equation}
    |\lambda_2|\leq 1 - (n-2)a_{\min} - 2\min\{a_{\min},c_{\min}\}\,.
\end{equation}
Using that the trace of $P$ satisfies
\begin{equation}
    n\cormin\leq \tr(P) = 1 + \sum_{i\neq 1}\lambda_i \leq  n\cormax\,,
\end{equation}
and the triangle inequality, separately for the left-hand and right-hand side of it, and combining each inequality into one, we have
\begin{equation}
    \max\{n\cormin - 1, 1-n\cormax ,0\} \leq |\sum_{i\neq 1}\lambda_i|   \leq  \sum_{i\neq 1} |\lambda_i| \leq (r-1)(1 - (n-2)a_{\min} - 2\min\{a_{\min},c_{\min}\}\,),
\end{equation}

where $r$ is the rank of $P$, which upper bounds the number of nonzero eigenvalues. In the last inequality we used $|\lambda_i|\leq\tau_1(P)$ for all $i\neq1$, and that at most $r-1$ of these eigenvalues are nonzero. Note that for the left-hand side, because it is possible to have a negative value for both $n\cormin -1$ and $1-n\cormax$ simultaneously ($\cormin < 1/n$, $\cormax >  1/n$), we use the fact that the right-hand side is always a positive value; hence the full expression of the left-hand side takes the form as provided. Now, since $d_C$ is lower bounded by the rank $r$, this implies that
\begin{equation}
    d_C\geq r\geq \frac{\max\{n\cormin - 1, 1-n\cormax ,0\} }{1 - (n-2)a_{\min} - 2\min\{a_{\min},c_{\min}\}\,}+1\,.
\end{equation}
\end{proof}

\section{Proof of Corollary~\ref{thm:dCbound}}\label{app:ProofIneq}

\dCboundCorollary*

\begin{proof}
For convenience, we will partially reproduce the proof of Theorem~\ref{thm:dCboundGeneral}.
We use again the ergodicity coefficient,
\begin{equation}
    \tau_1(P):=1-\min_{j\neq k}\sum_i\min\{P_{ij},P_{ik}\}.
\end{equation}
 This non-negative quantity  bounds  every eigenvalue $\lambda\neq1$ of $P$, $|\lambda|\leq \tau_1(P)$. In addition,
 if $r=\rank(P)$, then
\begin{equation}
    |\tr(P)-1|=\left|\sum_{\lambda\neq1}\lambda\right|\leq (r-1)\tau_1(P).
\end{equation}
Thus, whenever $\tau_1(P)>0$,
\begin{equation}\label{eq:masterRankBound}
    d_C\geq \rank(P)\geq 1+\frac{|\tr(P)-1|}{\tau_1(P)} .
\end{equation}
If $\tau_1(P)=0$, all columns of $P$ coincide, so $\rank(P)=1$.

We first consider the correlated case. Since $c_{\min}\geq a_{\max}$, all entries of $P$ are at least $a_{\min}$, and all diagonal entries are at least $c_{\min}$. Hence
\begin{equation}
    \tr(P)\geq n c_{\min},
    \qquad
    \sum_i\min\{P_{ij},P_{ik}\}\geq n a_{\min}
    \quad (j\neq k).
\end{equation}
Therefore $\tau_1(P)\leq 1-na_{\min}$. If $1-na_{\min}=0$, then all entries of $P$ are equal to $1/n$, so $P=J/n$ and $d_C=1$. Otherwise, using $|\tr(P)-1|\geq n c_{\min}-1$ and \eqref{eq:masterRankBound},
\begin{equation}
    d_C\geq 1+\frac{|\tr(P)-1|}{\tau_1(P)}
    \geq 1+\frac{n c_{\min}-1}{1-n a_{\min}},
\end{equation}

We now consider the anti-correlated case. Since $c_{\max}\leq a_{\min}$, every diagonal entry is smaller than or equal to every off-diagonal entry. In particular, choose a column $j^{*}$ such that $P_{j^{*}j^{*}}=c_{\max}$. Then, by column-stochasticity,
\begin{equation}
    1=\sum_i P_{ij^{*}}\geq n c_{\max},
\end{equation}
and hence $c_{\max}\leq 1/n$. Therefore
\begin{equation}
    1-nc_{\max}\geq0.
\end{equation}
In addition, trivially $\tr(P)\leq n c_{\max}$.

For any two distinct columns $j\neq k$, the two rows $i=j,k$ contribute at least $c_{\min}$ each, while the remaining $n-2$ rows contribute at least $a_{\min}$. Thus
\begin{equation}
    \sum_i\min\{P_{ij},P_{ik}\}\geq 2c_{\min}+(n-2)a_{\min},
\end{equation}
and therefore
\begin{equation}
    \tau_1(P)\leq 1-2c_{\min}-(n-2)a_{\min}.
\end{equation}

Again, if the denominator vanishes, then the inequalities above force $c_{\min}=a_{\min}=1/n$ and, since $c_{\max}\leq a_{\min}$, also $c_{\max}=1/n$. Hence $P=J/n$ and $d_C=1$. Otherwise, Eq.~\eqref{eq:masterRankBound} gives
\begin{equation}
    d_C\geq 1+\frac{|\tr(P)-1|}{\tau_1(P)}
    \geq 1+\frac{1-n c_{\max}}{1-2c_{\min}-(n-2)a_{\min}}.
\end{equation}
Finally, since $c_{\min}\geq0$,
\begin{equation}
    1-2c_{\min}-(n-2)a_{\min}\leq 1-(n-2)a_{\min}.
\end{equation}
Together with $1-nc_{\max}\geq0$ and positive denominators (i.e., $\tau_1(P)>0$), this also gives the weaker bound $d_C\geq 1+(1-nc_{\max})/(1-(n-2)a_{\min})$.
\end{proof}

\section{Stability of quadratic quantum advantage in $P_\alpha$ under experimental noise}\label{app:RankStabilityPalpha}
We will consider $E=P_\alpha-P'$, the matrix of experimental errors. The largest singular value of $E$ is equal to the spectral norm, $\|E\|_2$. In turn, by definition $\|E\|_2$ is larger or equal than any other singular value $\sigma_r(E)$ of $E$. Sorting the singular values of $E$ non-increasingly, $\sigma_1(E)=\|E\|_2\geq\dots\geq\sigma_n(E)\geq0$, it follows from Weyl's inequalities for singular values~\cite{HornJohnson2012} that
\begin{equation}
    |\sigma_r(P_\alpha)-\sigma_r(P')|\leq \|E\|_2
\end{equation}
for any pair of $r$'th singular values of $P_\alpha$ and $P'$. Therefore the condition $\|E\|_2<\sigma_r(P_\alpha)$ ensures that $\sigma_r(P')$ is nonzero and therefore $P'$ has rank at least $r$. In particular, $P'$ is full rank as long as
\begin{equation}
    \|E\|_2<\sigma_n(P_\alpha)=\frac{|n\alpha-1|}{n-1}\,.
\end{equation}

Here we used that $P_\alpha$ is symmetric, so that its singular values are $1$ and $|n\alpha-1|/(n-1)$. If $|E_{ij}|\leq\varepsilon$ for all $i,j$, then $\|E\|_2\leq\|E\|_F\leq n\varepsilon$, so that $\varepsilon<|n\alpha-1|/(n(n-1))$ suffices.

Alternatively, Theorem~\ref{thm:dCboundGeneral} yields robustness using only the extremal entries of $P'$. Let $|E_{ij}|\leq\varepsilon$ and $a:=(1-\alpha)/(n-1)$, so that $1-na=(n\alpha-1)/(n-1)$. For $\alpha>1/n$ one has $a<\alpha$, $c'_{\min}\geq\alpha-\varepsilon$ and $\min\{a'_{\min},c'_{\min}\}\geq a-\varepsilon$, hence
\begin{equation}
    1+\Gamma(P')\geq 1+\frac{n(\alpha-\varepsilon)-1}{1-n(a-\varepsilon)},
\end{equation}
which exceeds $n-1$ if and only if $\varepsilon<(n\alpha-1)/(n(n-1)^2)$. For $\alpha<1/n$ one has $\alpha<a$, $c'_{\max}\leq\alpha+\varepsilon$ and $\min\{a'_{\min},c'_{\min}\}\geq\alpha-\varepsilon$, which gives $1+\Gamma(P')\geq1+\frac{1-n(\alpha+\varepsilon)}{(1-n\alpha)/(n-1)+n\varepsilon}$ and the same condition with $1-n\alpha$. Since $d_C$ is an integer, $1+\Gamma(P')>n-1$ implies $d_C=n$.

\detBound*
\begin{proof}
Let $\{E_a\}_{a=1}^{n}$ be an orthonormal basis of the real space of Hermitian operators on $\C^d$, $\tr(E_aE_b)=\delta_{ab}$, which has dimension $n=d^2$. Every Hermitian operator $X$ has coordinates $x_a=\tr(XE_a)$, and the Hilbert--Schmidt product is the dot product of coordinates, $\tr(XY)=\sum_ax_ay_a$. Hence, if $m_i,r_j\in\R^n$ denote the coordinate vectors of $M_i$ and $\varrho_j$, then
\begin{equation}
    P_{ij}=\tr(M_i\varrho_j)=m_i\cdot r_j,\qquad\text{i.e.,}\qquad P=AB,
\end{equation}
with the $n\times n$ matrices $A$ (rows $m_i$) and $B$ (columns $r_j$). Since $AA^T$ and $B^TB$ are the Gram matrices $G^M_{ik}=\tr(M_iM_k)$ and $G^\varrho_{jk}=\tr(\varrho_j\varrho_k)$,
\begin{equation}\label{eq:detGram}
    (\det P)^2=(\det A)^2(\det B)^2=\det(G^M)\det(G^\varrho).
\end{equation}

\emph{Gram determinant of $n$ states.} We claim that for any $n=d^2$ states $\sigma_1,\dots,\sigma_n$ on $\C^d$, the Gram matrix $G_{jk}=\tr(\sigma_j\sigma_k)$ satisfies
\begin{equation}\label{eq:detGramStates}
    \det G\leq d\,\Big(\frac{d}{d+1}\Big)^{n-1}.
\end{equation}
Let $\mu_1\geq\dots\geq\mu_n\geq0$ be the eigenvalues of $G$. The largest eigenvalue is at least $u^TGu$ for any unit vector $u$; choosing $u=\mathbf 1/\sqrt n$,
\begin{equation}\label{eq:mu1}
    \mu_1\geq\frac1n\sum_{j,k}\tr(\sigma_j\sigma_k)=n\,\tr(\bar\sigma^2)\geq\frac nd,
    \qquad \bar\sigma=\frac1n\sum_j\sigma_j,
\end{equation}
where the last step uses that $\bar\sigma$ is a state on $\C^d$. The remaining eigenvalues satisfy
\begin{equation}
    \sum_{k\geq2}\mu_k=\tr G-\mu_1=\sum_j\tr(\sigma_j^2)-\mu_1\leq n-\mu_1,
\end{equation}
and, since the geometric mean of nonnegative numbers is upper bounded by their arithmetic mean,
\begin{equation}
    \prod_{k\geq2}\mu_k\leq\Big(\frac{1}{n-1}\sum_{k\geq2}\mu_k\Big)^{n-1}\leq\Big(\frac{n-\mu_1}{n-1}\Big)^{n-1}.
\end{equation}
Therefore
\begin{equation}\label{eq:detGmu1}
    \det G=\mu_1\prod_{k\geq2}\mu_k\leq\mu_1\Big(\frac{n-\mu_1}{n-1}\Big)^{n-1}.
\end{equation}
The right-hand side is decreasing in $\mu_1$ for $\mu_1\geq1$, so by Eq.~\eqref{eq:mu1} we may replace $\mu_1$ by $n/d$, which gives Eq.~\eqref{eq:detGramStates}. Equality holds if and only if all $\sigma_j$ are pure, $\bar\sigma=\one/d$ and $\mu_2=\dots=\mu_n$, i.e., $G=a\one+bJ$: the states form a SIC.

\emph{Gram determinant of the effects.} Write $M_i=t_i\sigma_i$ with $t_i=\tr(M_i)$ and $\sigma_i$ a state (if some $t_i=0$ then $\det P=0$ and there is nothing to prove). Then $G^M=\diag(t)\,G^\sigma\diag(t)$, so that
\begin{equation}
    \det G^M=\prod_it_i^2\,\det G^\sigma .
\end{equation}
Using again that the geometric mean is upper bounded by the arithmetic mean, and $\sum_it_i=\tr\one=d$,
\begin{equation}
    \prod_it_i\leq\Big(\frac1n\sum_it_i\Big)^n=\Big(\frac dn\Big)^n ,
\end{equation}
which together with Eq.~\eqref{eq:detGramStates} gives $\det G^M\leq(d/n)^{2n}\,d\,(d/(d+1))^{n-1}$.

\emph{Conclusion.} Inserting both bounds in Eq.~\eqref{eq:detGram} and using $n=d^2$,
\begin{equation}
    (\det P)^2\leq d^{-2n}\,d^2\Big(\frac{d}{d+1}\Big)^{2(n-1)}=(d+1)^{-2(n-1)}.
\end{equation}
Since $\lambda_1=1$ and $\prod_i\lambda_i=\det P$,
\begin{equation}
    |\lambda_{\min}|^{n-1}\leq\prod_{i\neq1}|\lambda_i|=|\det P|\leq(d+1)^{-(n-1)},
\end{equation}
and since $\sigma_1=\|P\|_2\geq\|P^T\mathbf 1\|/\|\mathbf 1\|=1$ for column-stochastic $P$, also $\sigma_{\min}^{n-1}\leq\prod_{i\neq1}\sigma_i\leq|\det P|$. For $P_\alpha$ with $\alpha=1/d$ the $n-1$ non-unit eigenvalues equal $1/(d+1)$, so all bounds are attained.
\end{proof}

\section{Maximal saturation selects $P_\alpha$ (proof of Lemma~\ref{lem:maxSaturationPalpha})}
\label{app:select_Palpha}

\maxSaturationPalpha*

\begin{proof}
Let $N:=\max\{nc_{\min}-1,\,1-nc_{\max},\,0\}$ and $D:=1-(n-2)a_{\min}-2\min\{a_{\min},c_{\min}\}>0$, so that the bound~\eqref{eq:UnifiedBound} reads $d_C\geq 1+N/D$, and it attains $1+N/D=n$ if and only if $N\geq(n-1)D$ (it cannot exceed $n$, since $1+N/D\leq d_C\leq n$). Since $P$ is column-stochastic, for every column $j$,
\begin{equation}
    1=\sum_iP_{ij}\ge P_{jj}+(n-1)a_{\min}\geq c_{\min}+(n-1)a_{\min}.
    \label{eq:normalization}
\end{equation}
If $N=0$, the bound cannot attain $n\geq 2$. Since $c_{\min}\leq c_{\max}$, at most one of $nc_{\min}-1$ and $1-nc_{\max}$ is positive, so two cases remain.

\noindent\emph{Case $N=nc_{\min}-1>0$ (correlated case).} If $c_{\min}<a_{\min}$, then all entries of $P$ are larger than $1/n$, which contradicts column-stochasticity. Hence $\min\{a_{\min},c_{\min}\}=a_{\min}$ and $D=1-na_{\min}$. The condition $N\geq(n-1)D$ then reads
\begin{equation}
    n c_{\min}-1 \ge (n-1)(1-n a_{\min}),
    \quad\text{i.e.,}\quad
    c_{\min}+(n-1)a_{\min}\ge1 .
\end{equation}
Together with Eq.~\eqref{eq:normalization}, $c_{\min}+(n-1)a_{\min}=1$, and equality holds termwise in every column: $P_{jj}=c_{\min}$ and $P_{ij}=a_{\min}$ for all $i\neq j$. Therefore $P=P_\alpha$ with $\alpha=c_{\min}>1/n$.

\noindent\emph{Case $N=1-nc_{\max}>0$ (anti-correlated case).} If $a_{\min}<c_{\min}$, then $D=1-na_{\min}>1-nc_{\max}=N>0$, so that $(n-1)D>N$, a contradiction. Hence $\min\{a_{\min},c_{\min}\}=c_{\min}$, and the condition $N\geq(n-1)D$ reads
\begin{equation}\label{eq:anti-lower-maxsat}
    (n-1)(n-2)a_{\min}\ge (n-2)-2(n-1)c_{\min}+n c_{\max}.
\end{equation}
On the other hand, Eq.~\eqref{eq:normalization} implies
\begin{equation}\label{eq:anti-upper-maxsat}
    (n-1)(n-2)a_{\min}\le(n-2)(1-c_{\min}).
\end{equation}
Combining~\eqref{eq:anti-lower-maxsat} and~\eqref{eq:anti-upper-maxsat} gives $n(c_{\max}-c_{\min})\le0$. Since $c_{\max}\ge c_{\min}$, we get $c_{\max}=c_{\min}$. Then~\eqref{eq:anti-lower-maxsat} and~\eqref{eq:anti-upper-maxsat} are tight, so that $c_{\min}+(n-1)a_{\min}=1$ for $n\geq3$. As in the correlated case, column-stochasticity forces $P_{jj}=c_{\min}$ and $P_{ij}=a_{\min}$ for all $i\neq j$. Thus $P=P_\alpha$ with $\alpha=c_{\min}=c_{\max}<1/n$.

\noindent\emph{Converse.} For $P_\alpha$ with $\alpha>1/n$, one has $c_{\min}=c_{\max}=\alpha$ and $a_{\min}=(1-\alpha)/(n-1)<\alpha$, so that $N=n\alpha-1$ and $D=(n\alpha-1)/(n-1)$. For $\alpha<1/n$, one has $a_{\min}>\alpha$, so that $N=1-n\alpha$ and $D=(1-n\alpha)/(n-1)$. In both cases $1+N/D=n$.
\end{proof}

\section{Proof of Proposition~\ref{prop:dQupper}}\label{app:dQupper}

\dQupper*

\begin{proof}
Let $d:=\lceil\sqrt{n}\rceil$, so that $d\geq2$ and $n\leq d^2$. We construct a single measurement with $n$ outcomes on $\C^d$, and for each column of $P$ a state reproducing it. We use three elementary facts:
\begin{enumerate}[label=(\roman*)]
    \item There are $d^2-1$ traceless Hermitian operators $G_1,\dots,G_{d^2-1}$ on $\C^d$ with $\tr(G_kG_l)=\delta_{kl}$, e.g., the normalized generalized Gell-Mann matrices. Since $n-1\leq d^2-1$, we use the first $n-1$ of them.
    \item There are vectors $s_1,\dots,s_n\in\R^{n-1}$ with $s_i\cdot s_k=\delta_{ik}-1/n$, namely the vertices of a regular simplex: the projections of the canonical basis of $\R^n$ onto the hyperplane orthogonal to $(1,\dots,1)$, written in an orthonormal basis of that hyperplane. They satisfy $\sum_is_i=0$, since $|\sum_is_i|^2=\sum_{i,k}(\delta_{ik}-1/n)=0$.
    \item Every traceless Hermitian $X$ on $\C^d$ satisfies $\|X\|_\infty\leq\kappa_d\|X\|_2$, with $\kappa_d:=\sqrt{(d-1)/d}$ and $\|X\|_2:=\sqrt{\tr(X^2)}$. Indeed, if $\lambda$ is the eigenvalue of $X$ with largest modulus, the remaining $d-1$ eigenvalues sum to $-\lambda$, so that their squares sum to at least $\lambda^2/(d-1)$, and $\tr(X^2)\geq\lambda^2 d/(d-1)$.
\end{enumerate}
Define $B_i:=\sum_{k=1}^{n-1}(s_i)_kG_k$. By (i) and (ii), $\tr(B_i)=0$, $\sum_iB_i=0$ and $\tr(B_iB_k)=s_i\cdot s_k=\delta_{ik}-1/n$. In particular $\|B_i\|_2^2=1-1/n$, and by (iii) $\|B_i\|_\infty\leq\kappa_d\sqrt{1-1/n}=:1/\tau$.

\emph{Measurement.} Let $M_i:=(\one+\tau B_i)/n$. Then $M_i\geq0$, since $\|\tau B_i\|_\infty\leq1$, and $\sum_iM_i=\one$.

\emph{States.} For each column $j$, let $\delta_i:=P_{ij}-1/n$, so that $\sum_i\delta_i=0$, and define
\begin{equation}
    \varrho_j:=\frac{\one}{d}+\frac{n}{\tau}X_j,\qquad X_j:=\sum_{i=1}^n\delta_iB_i .
\end{equation}
Then $\varrho_j$ is Hermitian with unit trace, and, using $\tr(B_k)=0$ and $\sum_i\delta_i=0$,
\begin{equation}
    \tr(\varrho_jM_k)=\frac1n+\tr(X_jB_k)=\frac1n+\sum_i\delta_i\Big(\delta_{ik}-\frac1n\Big)=\frac1n+\delta_k=P_{kj}.
\end{equation}
It remains to show $\varrho_j\geq0$. Since $\|X_j\|_2^2=\sum_{i,k}\delta_i\delta_k(\delta_{ik}-1/n)=\sum_i\delta_i^2$, fact (iii) gives
\begin{equation}
    \Big\|\frac{n}{\tau}X_j\Big\|_\infty\leq\frac{n\kappa_d}{\tau}\Big(\sum_i\delta_i^2\Big)^{1/2}=\frac{(d-1)\sqrt{n(n-1)}}{d}\Big(\sum_i\delta_i^2\Big)^{1/2}.
\end{equation}
Hence $\varrho_j\geq0$ whenever $\sum_i\delta_i^2\leq1/[(d-1)^2n(n-1)]$. Finally,
\begin{equation}
    \sum_i\delta_i^2=\Big(P_{jj}-\frac1n\Big)^2+\sum_{i\neq j}\Big(P_{ij}-\frac1n\Big)^2\leq\delta_c^2+(n-1)\,\delta_a^2 ,
\end{equation}
so that Eq.~\eqref{eq:dQupper} implies $\varrho_j\geq0$ for all $j$. Therefore the states $\{\varrho_j\}$ and the measurement $\{M_i\}$ reproduce $P$ in dimension $d$.
\end{proof}

\emph{SIC refinement.} If $n=d^2$ and a SIC-POVM $\{\Pi_i/d\}$ exists in dimension $d$, the same strategy with $M_i=\Pi_i/d$ and
\begin{equation}
    \varrho_j:=(d+1)\sum_iP_{ij}\Pi_i-\one
\end{equation}
gives a complementary region. Indeed, $\tr(\varrho_j)=1$, and using $\tr(\Pi_i\Pi_k)=(d\delta_{ik}+1)/(d+1)$ and $\sum_iP_{ij}=1$ one finds $\tr(\varrho_jM_k)=[dP_{kj}+1-1]/d=P_{kj}$. Moreover, since $\sum_i\Pi_i=d\one$, $P_{ij}\geq a_{\min}$ for $i\neq j$ and $0\leq\Pi_j\leq\one$,
\begin{equation}
    \sum_iP_{ij}\Pi_i\geq P_{jj}\Pi_j+a_{\min}(d\one-\Pi_j)\geq\big[(d-1)a_{\min}+\min\{a_{\min},c_{\min}\}\big]\one .
\end{equation}
Hence $\varrho_j\geq0$ whenever $(d-1)a_{\min}+\min\{a_{\min},c_{\min}\}\geq1/(d+1)$, with equality for both $P_\alpha$ with $\alpha=1/d$ (optimal discrimination) and $\alpha=0$ (optimal exclusion).

\section{Proof of Theorem~\ref{thm:selftest}}\label{app:SelfTest}

\selftestSIC*

\begin{proof}
If the right-hand side of Eq.~\eqref{eq:dCboundDisc} equals $n$ with $c_{\min}=1/d_Q$, then $1-na_{\min}=(n/d_Q-1)/(n-1)=1/(d_Q+1)$, i.e., $a_{\min}=1/[d_Q(d_Q+1)]$, and every column satisfies $1=\sum_iP_{ij}\geq c_{\min}+(n-1)a_{\min}=1$. Hence equality holds termwise, $P_{jj}=1/d_Q$ and $P_{ij}=a_{\min}$ for $i\neq j$, so that $P=P_\alpha$ with $\alpha=1/d_Q=1/\sqrt{n}$. Conversely, SIC-POVMs with aligned states produce this $P_\alpha$. Hence it suffices to prove that the only possible quantum strategy to obtain $P_\alpha$ with $\alpha=1/d_Q$ requires using SIC-POVMs. For brevity we write $d:=d_Q$.
We start from the diagonal entries of the CM:
\begin{equation}
   p(j|j)= \frac{1}{d}=\tr(\varrho_jM_j)\leq\tr(M_j)
\end{equation}
which implies that $\tr(M_j)\geq 1/d$. On the other hand, $\sum_iM_i=\one$ and therefore $\sum_{i=1}^{d^2}\tr(M_i)=d$, which implies that the inequality is tight, $\tr(M_j)=1/d$. Now, using the Cauchy--Schwarz inequality we have that
\begin{equation}\label{eq:UppLowBoundTrM}
    \frac{1}{d}=\tr(\varrho_iM_i)\leq\sqrt{\tr(\varrho_i^2)\tr(M_i^2)}\leq\sqrt{\tr(M_i^2)}\leq\sqrt{\tr(M_i)^2}=\tr(M_i)=\frac{1}{d}.
\end{equation}
Here the second inequality uses $\tr(\varrho_i^2)\le 1$ for density operators; and the third uses positivity of $M_i$, since in its spectral decomposition $\tr(M_i^2)=\sum_k\lambda_k^2 \le (\sum_k\lambda_k)^2 =\tr(M_i)^2$. As the left- and right-hand sides of~\eqref{eq:UppLowBoundTrM} coincide, all inequalities are saturated. Hence $\varrho_i$ is pure and equality in Cauchy--Schwarz gives $M_i\propto \varrho_i$. Together with $\tr(M_i)=1/d$, this yields $\varrho_i=\Pi_i=\dyad{\psi_i}$ and $M_i=\dyad{\psi_i}/d$.
Finally, the SIC overlaps $\tr(\Pi_i \Pi_j)=\frac{1}{d+1}$ for $i\neq j$ are recovered from the off-diagonal entries of the CM, $p(i|j)=\tr(M_i \varrho_j)=\frac{1}{d(d+1)}$.
\end{proof}

\section{Proofs for robustness of SIC-certification}\label{app:robustness}

\begin{proposition}[Continuity of Born probabilities]\label{prop:continuity}
Assume that the ideal target realization $\{\varrho_j,M_i\}$ acts on a $d_Q$-dimensional Hilbert space $\mathcal H$, and that the true noisy experimental realization $\{\tilde\varrho_j,\tilde M_i\}$ acts on a possibly larger Hilbert space $\mathcal H_{\rm exp}\supseteq\mathcal H$. With a slight abuse of notation, we identify the ideal operators with their trivial extensions to $\mathcal H_{\rm exp}$ and define the implementation errors
\begin{equation}
    \Delta\varrho_j:=\tilde\varrho_j-\varrho_j,
    \qquad
    \Delta M_i:=\tilde M_i-M_i .
\end{equation}
Then, for all $i,j$,
\begin{equation}
    \left|\tr(\tilde\varrho_j\tilde M_i)-\tr(\varrho_jM_i)\right|
    \le
    \|\Delta\varrho_j\|_1 \|M_i\|_\infty + \|\Delta M_i\|_\infty .
\end{equation}
\end{proposition}
In particular, since $0\le M_i\le\one$,
\begin{equation}
    \left|\tr(\tilde\varrho_j\tilde M_i)-\tr(\varrho_jM_i)\right|
    \le
    \|\Delta\varrho_j\|_1 + \|\Delta M_i\|_\infty .
\end{equation}
For the target of Theorem~\ref{thm:selftest}, namely SIC-POVMs, where $M_i=\dyad{\psi_i}/d_Q$, one has $\|M_i\|_\infty=1/d_Q$, and therefore
\begin{equation}
    \left|\tr(\tilde\varrho_j\tilde M_i)-\tr(\varrho_jM_i)\right|
    \le
    \frac{1}{d_Q}\|\Delta\varrho_j\|_1 + \|\Delta M_i\|_\infty .
\end{equation}

\begin{proof}[Proof of Proposition~\ref{prop:continuity}]
With a slight abuse of notation, we use $\varrho_j$ and $M_i$ also for their trivial extensions $\varrho_j\oplus 0$ and $M_i\oplus 0$ to $\mathcal H_{\rm exp}$. In particular $\tr(\varrho_j M_i)$
is unchanged when computed on $\mathcal H_{\rm exp}$. Hence
\begin{align}
\tr(\tilde\varrho_j\tilde M_i)-\tr(\varrho_jM_i)
&=
\tr\bigl(\tilde\varrho_j(M_i+\Delta M_i)\bigr)
-\tr(\varrho_jM_i)
=
\tr(\Delta\varrho_j M_i)
+
\tr(\tilde\varrho_j\Delta M_i).
\end{align}
By H\"older's inequality,
\begin{equation}
    \left|\tr(\Delta\varrho_jM_i)\right|
    \le
    \|\Delta\varrho_j\|_1\|M_i\|_{\infty}.
\end{equation}
Moreover, since $\tilde\varrho_j$ is a density operator,
\begin{equation}
    \left|\tr(\tilde\varrho_j\Delta M_i)\right|
    \le
    \|\Delta M_i\|_{\infty}.
\end{equation}
Combining both estimates gives
\begin{equation}
    \left|
    \tr(\tilde\varrho_j\tilde M_i)
    -
    \tr(\varrho_jM_i)
    \right|
    \le
    \|\Delta\varrho_j\|_1\|M_i\|_{\infty}
    +
    \|\Delta M_i\|_{\infty}.
\end{equation}
Since $0\le M_i\le \one$, one has $\|M_i\|_{\infty}\le 1$, which gives the general bound. For the SIC target of Theorem~\ref{thm:selftest}, $M_i=\dyad{\psi_i}/d_Q$, and therefore $\|M_i\|_{\infty}=1/d_Q$. Hence
\begin{equation}
   |\tr(\tilde\varrho_j\tilde M_i)
    -\tr(\varrho_jM_i)|
    \le
    \frac{1}{d_Q}\|\Delta\varrho_j\|_1
    +  \|\Delta M_i\|_{\infty}.
\end{equation}
\end{proof}

\robustSIC*

\begin{proof}[Proof of Theorem~\ref{thm:robustSIC}]
For the ideal SIC communication matrix,
\begin{equation}
    P^{\rm SIC}_{ii}=\frac1{d_Q},
    \qquad
    P^{\rm SIC}_{ij}=\frac1{d_Q(d_Q+1)}
    \quad (i\neq j).
\end{equation}
Assume
\begin{equation} \label{eq:robust-assumption}
    |P_{ij}-P^{\rm SIC}_{ij}| \le \varepsilon
    \qquad
    \forall i,j .
\end{equation}

For the diagonal elements, \eqref{eq:robust-assumption} gives
\begin{equation}\label{eq:Pii-bounds}
    \frac1{d_Q}-\varepsilon \le P_{ii} \le \frac1{d_Q}+\varepsilon .
\end{equation}

Since $\varrho_i\le\one$,
\begin{equation}
    P_{ii} = \tr(\varrho_i M_i) \le \tr(M_i),
\end{equation}
and therefore
\begin{equation}\label{eq:trace-lower}
    \tr(M_i) \ge \frac1{d_Q}-\varepsilon .
\end{equation}

Next, because $\sum_i M_i=\one$,
\begin{equation}
    \sum_{i=1}^{d_Q^2}\tr(M_i)=\tr(\one)=d_Q .
\end{equation}
Using \eqref{eq:trace-lower} for all $k\neq i$,
\begin{align}
    \tr(M_i)
    &= d_Q-\sum_{k\neq i}\tr(M_k)\\
    &\le d_Q-(d_Q^2-1)\left(\frac1{d_Q}-\varepsilon\right)= \frac1{d_Q}+(d_Q^2-1)\varepsilon .
\end{align}
Hence
\begin{equation}\label{eq:trace-bounds}
    \frac1{d_Q}-\varepsilon
    \le
    \tr(M_i)
    \le
    \frac1{d_Q}+(d_Q^2-1)\varepsilon .
\end{equation}

Let $\lambda_i:=\|M_i\|_\infty$ denote the largest eigenvalue of $M_i$. Since $M_i\ge0$, $P_{ii}=\tr(\varrho_i M_i)\le\lambda_i$,
while trivially
\begin{equation}
    \lambda_i\le\tr(M_i).
\end{equation}
Combining this with equations \eqref{eq:Pii-bounds} and \eqref{eq:trace-bounds} gives
\begin{equation}\label{eq:lambda-bounds}
    \frac1{d_Q}-\varepsilon
    \le
    \lambda_i
    \le
    \frac1{d_Q}+(d_Q^2-1)\varepsilon,
\end{equation}
which in particular implies
\begin{equation}\label{eq:lambda-abs}
    \left|\lambda_i-\frac1{d_Q}\right|
    \le
    (d_Q^2-1)\varepsilon .
\end{equation}

Furthermore, since $\lambda_i\ge P_{ii}$,
\begin{align}
\label{eq:trM-lambda}
    \tr(M_i)-\lambda_i
    &\le \tr(M_i)-P_{ii}\nonumber\\
    &\le
    \left(\frac1{d_Q}+(d_Q^2-1)\varepsilon\right)
    -
    \left(\frac1{d_Q}-\varepsilon\right)= d_Q^2\varepsilon .
\end{align}
where in the second inequality we used   \eqref{eq:Pii-bounds} and \eqref{eq:trace-bounds}.

Let $\Pi_i$ be a rank-one projector onto an eigenvector associated to $\lambda_i$. By the spectral decomposition,
\begin{equation}
    M_i=\lambda_i\Pi_i+R_i ,
\end{equation}
where $R_i\ge0$ and $R_i\Pi_i=0$. Since $(\lambda_i-1/d_Q)\Pi_i$ and $R_i$ have orthogonal supports,
\begin{align}
    \left\|M_i-\frac1{d_Q}\Pi_i\right\|_\infty
    &= \max\left\{\left|\lambda_i-\frac1{d_Q}\right|,\|R_i\|_\infty\right\}
    \le \max\left\{\left|\lambda_i-\frac1{d_Q}\right|,\tr(R_i)\right\}.
\end{align}
The first term is bounded by~\eqref{eq:lambda-abs}, and the second by~\eqref{eq:trM-lambda}, since $\tr(R_i)=\tr(M_i)-\lambda_i\le d_Q^2\varepsilon$. Therefore
\begin{equation}\label{eq:Mi-close}
    \left\|M_i-\frac1{d_Q}\Pi_i\right\|_\infty
    \le
    d_Q^2\varepsilon .
\end{equation}

Next, we write the diagonal element of the CM as
\begin{equation}
  \frac{1}{d_Q}  \tr(\varrho_i\Pi_i)=
    P_{ii}
    -
    \tr\!\left[\varrho_i\left(M_i-\frac1{d_Q}\Pi_i\right)\right]
    \ge
     \frac{1}{d_Q} -(d_Q^2+1)\varepsilon ,
\end{equation}
where we used $P_{ii}\ge1/d_Q-\varepsilon$ and $|\tr[\varrho_i(M_i-\Pi_i/d_Q)]|\le\|M_i-\Pi_i/d_Q\|_\infty$ together with~\eqref{eq:Mi-close}. Multiplying by $d_Q$ gives
\begin{equation}
    \tr(\varrho_i\Pi_i)
    \ge
    1-d_Q(d_Q^2+1)\varepsilon .
\end{equation}
By the Fuchs--van de Graaf inequality,
\begin{equation}\label{eq:rho-close}
    \|\varrho_i-\Pi_i\|_1
    \le
    2\sqrt{1-\tr(\varrho_i\Pi_i)}\leq
    2\sqrt{d_Q(d_Q^2+1)\varepsilon}.
\end{equation}

Finally, for the off-diagonal entries $P_{ji}=\tr(\varrho_iM_j)$, the triangle inequality gives
\begin{align}
    \left|
    P_{ji}-\frac1{d_Q}\tr(\Pi_i\Pi_j)
    \right|
    &\le
    \left|
    \tr\left[\varrho_i\left(M_j-\frac1{d_Q}\Pi_j\right)\right]
    \right|+
    \frac1{d_Q}\left|
   \tr\left[(\varrho_i-\Pi_i)\Pi_j\right]
    \right|\nonumber\\
    &\le
    \left\|M_j-\frac1{d_Q}\Pi_j\right\|_\infty
    +
    \frac1{d_Q}\|\varrho_i-\Pi_i\|_1
    \le
    d_Q^2\varepsilon
    +
    \frac{2}{d_Q}\sqrt{d_Q(d_Q^2+1)\varepsilon}.
\end{align}
Using $|P_{ji}-P^{\rm SIC}_{ji}|\le\varepsilon$ and multiplying by $d_Q$, we obtain
\begin{equation}
    \left|
    \tr(\Pi_i\Pi_j)-\frac{d_Q\delta_{ij}+1}{d_Q+1}
    \right|
    \le
    d_Q(d_Q^2+1)\varepsilon
    +
    2\sqrt{d_Q(d_Q^2+1)\varepsilon},
\end{equation}
which trivially also holds for $i=j$. This concludes the proof.
\end{proof}

\begin{proposition}[Leakage robustness]\label{prop:leakage}
Assume the setting of Theorem~\ref{thm:robustSIC}, except that the experimental realization acts on a larger Hilbert space $\mathcal H_{\rm exp}\supseteq\mathcal H$. Let $P_{\mathcal H}$ denote the projector onto the intended $d_Q$-dimensional target subspace $\mathcal H$, and suppose that the preparations satisfy the leakage bound
\begin{equation}
    \tr\!\left[(\one-P_{\mathcal H})\tilde\varrho_i\right] \le \delta
    \qquad
    \forall i .
\end{equation}
Then the compressed realization on $\mathcal H$ satisfies the conclusions of Theorem~\ref{thm:robustSIC} with effective error parameter
\begin{equation}
    \varepsilon_{\rm eff} = \varepsilon + 2\sqrt{\delta} + 2\delta .
\end{equation}
\end{proposition}

\begin{proof}[Proof of Proposition~\ref{prop:leakage}]
Define the compressed states
\begin{equation}
    \widehat\varrho_i
    :=
    \frac{P_{\mathcal H}\tilde\varrho_iP_{\mathcal H}}
    {\tr(P_{\mathcal H}\tilde\varrho_i)},
\end{equation}
and compressed POVM elements
\begin{equation}
    \widehat M_i
    :=
    P_{\mathcal H}\tilde M_iP_{\mathcal H}.
\end{equation}
Write $Q:=\one-P_{\mathcal H}$ and $t_i:=\tr(P_{\mathcal H}\tilde\varrho_i)\geq1-\delta$. Decomposing $\tilde\varrho_i$ into blocks,
\begin{equation}
    \tr(\tilde\varrho_i\tilde M_j)
    =\tr(P_{\mathcal H}\tilde\varrho_iP_{\mathcal H}\tilde M_j)
    +\tr(Q\tilde\varrho_iQ\tilde M_j)
    +2\,{\rm Re}\,\tr(P_{\mathcal H}\tilde\varrho_iQ\tilde M_j).
\end{equation}
Since $0\le\tilde M_j\le\one$, the second term is at most $\tr(Q\tilde\varrho_i)\le\delta$. By the Cauchy--Schwarz inequality, $|\tr(P_{\mathcal H}\tilde\varrho_iQ\tilde M_j)|\le\sqrt{\tr(Q\tilde\varrho_i)}\sqrt{\tr(\tilde\varrho_iP_{\mathcal H}\tilde M_j^2P_{\mathcal H})}\le\sqrt{\delta}$. Moreover, $\tr(\widehat\varrho_i\widehat M_j)=\tr(P_{\mathcal H}\tilde\varrho_iP_{\mathcal H}\tilde M_j)/t_i$ and $0\le\tr(P_{\mathcal H}\tilde\varrho_iP_{\mathcal H}\tilde M_j)\le t_i$, so that normalization changes this term by at most $1-t_i\le\delta$. Altogether,
\begin{equation}
    \left| \tr(\tilde\varrho_i\tilde M_j)
    -  \tr(\widehat\varrho_i\widehat M_j) \right|
    \le
    2\sqrt{\delta}+2\delta .
\end{equation}
Therefore the compressed realization, which acts on the $d_Q$-dimensional space $\mathcal H$ with $\sum_j\widehat M_j=P_{\mathcal H}$, satisfies
\begin{equation}
    \left|\tr(\widehat\varrho_i\widehat M_j)-P^{\rm SIC}_{ji}\right|
    \le
    \varepsilon + 2\sqrt{\delta}+2\delta .
\end{equation}
Applying Theorem~\ref{thm:robustSIC} to the compressed realization proves the claim.
\end{proof}

\end{widetext}

\end{document}